\documentclass[particles,review,accept,pdftex,moreauthors]{Definitions/mdpi} 

\usepackage{bm}
\allowdisplaybreaks[4]
\firstpage{515} 
\pubvolume{6}
\issuenum{2}
\articlenumber{29}
\pubyear{2023}
\copyrightyear{2023}
\externaleditor{\hl{Academic Editor: } 
Firstname Lastname}
\datereceived{6 March 2023} 
\daterevised{2 April 2023} 
\dateaccepted{12 April 2023} 
\datepublished{} 
\hreflink{https://doi.org/10.3390/particles6020029} 

\Title{Several Topics on Transverse Momentum-Dependent Fragmentation Functions}

\TitleCitation{Several Topics on Transverse Momentum-Dependent Fragmentation Functions}

\Author{{Kai-Bao Chen} 
 $^{1,}$*\orcidA{}, {Tianbo Liu} $^{2,}$*\orcidB{}, Yu-Kun Song $^{3,}$*\orcidC{} and Shu-Yi Wei $^{2,}$*\orcidD{}} 

\AuthorNames{Kai-Bao Chen, Tianbo Liu, Yu-Kun Song and Shu-Yi Wei}

\AuthorCitation{Chen, K.-B.; Liu, T.; \linebreak Song, Y.-K.; Wei, S.-Y.}

\address{%
$^{1}$ \quad School of Science, Shandong Jianzhu University, {Jinan 250101,} 
 China\\
$^{2}$ \quad Key Laboratory of Particle Physics and Particle Irradiation (MOE), Institute of Frontier and \linebreak Interdisciplinary Science, Shandong University, Qingdao 266237, China\\
$^{3}$ \quad School of Physics and Technology, University of Jinan, Jinan 250022, China
}

\corres{Correspondence: chenkaibao19@sdjzu.edu.cn (K.-B.C.); liutb@sdu.edu.cn (T.L.); \linebreak  sps\_songyk@ujn.edu.cn (Y.-K.S.); shuyi@sdu.edu.cn (S.-Y.W.)}

\abstract{The hadronization of a high-energy parton is described by fragmentation functions which are introduced through QCD factorizations. While the hadronization mechanism per se remains uknown, fragmentation functions can still be investigated qualitatively and quantitatively. The qualitative study mainly concentrates on extracting genuine features based on the operator definition in quantum field theory. The quantitative research focuses on describing a variety of experimental data employing the fragmentation function given by the parameterizations or model calculations. With the foundation of the transverse-momentum-dependent factorization, the QCD evolution of leading twist transverse-momentum-dependent fragmentation functions has also been established. In addition, the universality of fragmentation functions has been proven, albeit model-dependently, so that it is possible to perform a global analysis of experimental data in different high-energy reactions. The collective efforts may eventually reveal important information hidden in the shadow of nonperturbative physics. This review covers the following topics: transverse-momentum-dependent factorization and the corresponding QCD evolution, spin-dependent fragmentation functions at leading and higher twists, several experimental measurements and corresponding phenomenological studies, and some model calculations. }

\keyword{fragmentation function; transverse-momentum-dependent factorization; QCD evolution; spin-related effects} 

\begin{document}

\section{Introduction}
\label{sec:intro}

Quantum chromodynamics (QCD)~\cite{Fritzsch:1973pi} is known as the fundamental theory of strong interaction in the framework of Yang-Mills gauge field theory~\cite{Yang:1954ek}. As a key property of QCD, the color confinement prohibits direct detection of quarks and gluons, the fundamental degrees of freedom, with any modern detectors. The emergence of color neutral hadrons from colored quarks and gluons is still an unresolved problem and has received particular interest in recent years~\cite{Metz:2016swz}. With the progress of QCD into the precision era, unraveling the hadronization mechanism in the high-energy scattering processes has become one of the most active frontiers in nuclear and particle physics.

Due to the nonperturbative nature of QCD, it is still challenging to directly calculate the hadronization process from first principles. Similar to the parton distribution functions (PDFs)~\cite{Bjorken:1969ja, Feynman:1973xc}, which were originally defined as the probability density of finding a parton inside the parent hadron, the concept of fragmentation functions (FFs) was introduced by Berman, Bjorken, and Kogut \cite{Berman:1971xz} right after the parton model to describe the emergence of a system of the hadron from a high-energy parton isolated in the phase space. An alternative name, the parton decay function, has also frequently been used in early literature. 

The modern concept of FFs in QCD was first introduced to describe the inclusive production of a desired hadron in the $e^+e^-$ annihilation~\cite{Mueller:1978xu, Collins:1981uw}, which is still the cleanest reaction currently available to investigate the fragmentation process. Within the QCD-improved parton model, the FF has its foundation in the factorization theorem~\cite{Collins:1989gx, Collins:2011zzd}, in which the differential cross section is approximated as a convolution of short-distance hard scattering and long-distance matrix elements with corrections formally suppressed by inverse powers of a hard scale, {e.g.,} the center-of-mass (c.m.) energy $Q=\sqrt{s}$ in the $e^+e^-$ annihilation. The predictive power of this theoretical framework relies on the control of the hard probe, which can be achieved by our ability to calculate the partonic cross section order by order in the perturbation theory, and the universality of the long-distance functions, such as the FFs, to be tested in multiple high-energy scattering processes. 

For a single-scale process, {e.g.,} $e^+e^- \to h X$, where $h$ represents the identified hadron in the final state and $X$ denotes the undetected particles, the process is not sensitive to the confined motion of quarks and gluons in the hadronization process, and one can apply the colinear factorization with the emergence of the detected hadron described by a colinear FF $D_{f\to h}(z)$, where the subscript $f$ stands for the parton flavor and $z$ is the longitudinal momentum fraction carried by the hadron $h$ with respect to the fragmenting parton. 
If two hadrons are identified in a process, {e.g.,} $e^+e^- \to h_A h_B X$, where $h_A$ and $h_B$ are detected hadrons in the final state, the reaction becomes a double-scale problem with one scale $Q$ given by the hard probe and the other scale provided by the transverse momentum imbalance, $|\bm{p}_{A\perp} + \bm{p}_{B\perp}|$. When the second scale is much smaller than $Q$, {i.e.,} the two hadrons are nearly back to back, one needs to use the transverse-momentum-dependent (TMD) factorization. The emergence of each of the hadrons is described by a TMD FF $D_{f\to h}(z,{k}_\perp)$, where ${k}_\perp$ is the transverse momentum of the fragmenting parton with respect to the observed hadron~\cite{Collins:1981uw,Collins:1981uk}. When the two scales are compatible, the reaction effectively becomes a single-scale process, and one can again use the colinear factorization. The matching between the two regions has been developed. The TMD FFs defined in the $e^+e^-$ annihilation also play an important role in the study of nucleon three-dimensional structures via the semi-inclusive deep inelastic scattering (SIDIS) process~\cite{Aybat:2011zv}. Instead of identifying two hadrons in a reaction, one can also access TMD FFs in the single-hadron production process by reconstructing the thrust axis, which provides the sensitivity to the transverse momentum of the observed hadron, as proposed in recent years~\cite{Kang:2020yqw,Boglione:2020auc,Boglione:2020cwn,Makris:2020ltr}.

Taking the parton spin degree of freedom into account, one can define polarized or spin-dependent TMD FFs. They essentially reflect the correlation between parton transverse momentum and its spin during the hadronization process and result in rich phenomena in high-energy scattering processes. For example, the Collins fragmentation function $H_1^\perp(z,{k}_\perp)$~\cite{Collins:1992kk}, naively interpreted as the probability density of a transversely polarized quark fragmenting into an unpolarized hadron, can lead to a single spin asymmetry (SSA) in the SIDIS process with a transversely polarized target~\cite{Boer:2003cm}. This asymmetry is a key observable for the determination of the quark transversity distribution, the net density of a transversely polarized quark in a transversely polarized nucleon. It also leads to azimuthal asymmetries in $e^+e^-$ annihilation as measured by Belle, BaBar, and BESIII. The progress of experimental techniques to determine the spin state of produced hyperons, such as $\Lambda$ and $\Omega$, and vector mesons, such as $\rho$ and $K^*$, offer us the opportunity to extract additional information from FFs. This is far beyond a trivial extension since the spin has been proven to be a powerful quantity to test theories and models, especially in hadron physics. The recent measurement of the spontaneous polarization of $\Lambda$ from unpolarized $e^+e^-$ annihilation is such an instance \cite{Belle:2018ttu}. This observation can be explained by a naively time-reversal odd (T-odd) TMD FF $D_{1T}^\perp(z,{k}_\perp)$ and has received interests from various groups \cite{Matevosyan:2018jht, Gamberg:2018fwy, Anselmino:2019cqd, Anselmino:2020vlp, DAlesio:2020wjq, Callos:2020qtu, Kang:2020xyq, Boglione:2020cwn, Li:2020oto, Chen:2021hdn, Gamberg:2021iat, DAlesio:2021dcx, DAlesio:2022brl,Boglione:2022nzq}.

In addition to the leading-twist FFs, which usually have probability interpretations, the high-twist FFs have been found o be much more important than expected in recent years for understanding precise experimental data \cite{Ellis:1982wd, Ellis:1982cd, Qiu:1988dn, Qiu:1990xxa, Qiu:1990xy, Balitsky:1990ck, Levelt:1994np, Levelt:1993ac, Kotzinian:1994dv, Mulders:1995dh, Boer:1997mf, Kotzinian:1997wt, Boer:1997nt, Boer:1997qn, Bacchetta:2000jk, Boer:1999uu, Bacchetta:2004zf, Bacchetta:2006tn, Boer:2008fr, Eguchi:2006qz, Eguchi:2006mc, Koike:2006qv, Kanazawa:2013uia, Pitonyak:2013dsu, Yang:2016qsf, Liang:2006wp, Liang:2008vz, Gao:2010mj, Song:2010pf, Song:2013sja, Wei:2013csa, Wei:2014pma, Chen:2016moq, Wei:2016far}. Although the colinear factorization at subleading power was demonstrated some time ago, the TMD factorization beyond the leading power is still under exploration, and some approaches have been proposed \cite{Vladimirov:2021hdn, Rodini:2022wki, Rodini:2022wic, Ebert:2021jhy, Balitsky:2017flc, Balitsky:2017gis, Balitsky:2020jzt, Balitsky:2021fer, Gamberg:2022lju}. Although high-twist contributions are formally power suppressed, their contributions to the cross section might not be negligible and may have significant effects in certain kinematics or observables. The inclusion of high-twist FFs will also modify the evolution equation and consequently affect the leading-twist FFs. The TMD factorization at subleading power was recently explored with different approaches. Overall, many efforts, both theoretical and experimental, are still required to understand the hadronization process and the upcoming data from future electron-ion colliders.

The remainder of this review is organized as follows. In Section \ref{sec:fac} we use $e^+e^- \to h_A h_B X$ as an example to present the flow of deriving the TMD factorization and the QCD evolution equation of TMD FFs. In Section \ref{sec:spin}, we present the FFs up to the twist-4 level for \linebreak  spin-$0$, -$1/2$, and $1$ hadron productions. In Section \ref{sec:exp}, we summarize the experimental measurements towards understanding the spin-dependent FFs. In Section \ref{sec:model}, we briefly lay out some model calculations. A summary is given in Section \ref{6}.

\section{Factorization and Evolution}
\label{sec:fac}

The modern concept of FFs has established on the QCD factorization theorems, which can be derived either from calculating traditional Feynman diagrams in perturbative field theory~\cite{Collins:1981uk, Collins:1981va, Collins:1984kg, Ji:2004wu, Ji:2004xq, Sun:2014gfa, Sun:2015doa, Collins:2011zzd} or in effective theories~\cite{Becher:2010tm,Echevarria:2011epo,Chiu:2012ir,Li:2016axz,Kang:2017glf}. In the former approach, one first identifies a collection of Feynman diagrams that offers the leading contribution through the Libby--Sterman analysis~\cite{Sterman:1978bi, Libby:1978bx}. In this method, the leading contribution is represented by the reduced diagrams. 

\begin{figure}[H]
\includegraphics[width=0.8\textwidth]{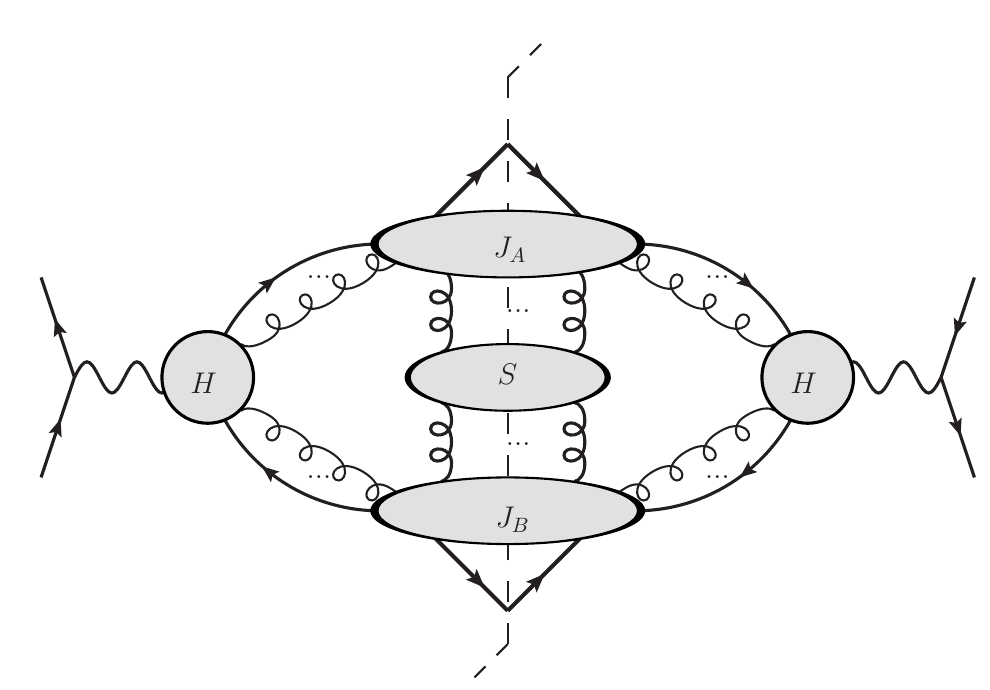}
\caption{Leading regions for $e^+e^-\to h_Ah_BX$.}
\label{fig:leading-region}
\end{figure}

Taking $e^+e^-\to h_Ah_BX$ process with $h_A,h_B$ traveling along almost back-to-back directions as an example~\cite{Collins:1981uk}, the leading regions are presented in Figure~\ref{fig:leading-region}. The cross section is the product of various ingredients, such as the hard part $H$, the soft part $S$, and the colinear parts $J_A$, $J_B$. We work in the light-cone coordinate, so that a four momentum $p$ can be written as follows: $p^\mu=(p^+,p^-,\bm{p}_\perp)$ with $p^\pm=\frac{1}{\sqrt{2}}(p^0\pm p^3)$. In the kinematic region where TMD factorization applies, the transverse momentum is considerably small compared with that along the longitudinal direction. Therefore, the momenta of the almost back-to-back hadrons A and B scale as $p_A\sim Q(1,\lambda,\sqrt{\lambda})$ and $p_B\sim Q(\lambda,1,\sqrt{\lambda})$, where $Q$ is the large momentum scale and $\lambda\ll 1$ is a small parameter. The hard part $H$ computes the cross section of interaction among hard partons whose momenta scale as $Q(1,1,1)$ in perturbative field theory. The contribution from colinear partons whose momenta are colinear with the final state hadrons A and B are evaluated in the colinear function $J_{A/B}$. This process results in the gauge invariant bare FFs. The soft part calculates the contribution from soft gluons whose momenta typically take the form of $Q(\lambda,\lambda,\lambda)$. They will be absorbed into the definition of TMD FFs eventually and convert the bare FFs into the \linebreak renormalized ones. 
\vspace{-12pt}

The interactions between different parts can be eliminated via applying appropriate kinematic approximations and the Ward identity. Finally, the cross section is given by a convolution of those well-separated parts, and we arrive at the factorization theorem of \linebreak this process. 

Depending on the physics of interest, we may derive either colinear factorization or transverse-momentum-dependent (TMD) factorization theorems. For the differential cross section of $e^+e^-\to h_Ah_BX$ as a function of the relative transverse momentum between $h_A$ and $h_B$, the TMD factorization theorem applies. 

In the single-photon-exchange approximation, the differential cross section of this process can be written as the production of a leptonic tensor and a hadronic tensor. \linebreak  It reads as follows:\cite{Boer:1997mf}
\begin{align}
\frac{d\sigma}{dy dz_A dz_B d^2 \bm{P}_{A\perp}} = \frac{2\pi N_c \alpha^2}{Q^4} L_{\mu\nu} W^{\mu\nu}, 
\end{align}
where, $\alpha$ is the coupling constant, $N_c=3$ is the color factor, $Q$ is the center-of-mass energy of the colliding leptons, $y=(1+\cos\theta)/2$ with $\theta$ the angle between incoming electron and  the outgoing hadron $h_A$, $z_A$ and $z_B$ are light-cone momentum fractions of $h_A$ and $h_B$, and $\boldsymbol{P}_{A\perp}$ is the transverse momentum of $h_A$ with respect to the direction of the $h_B$ momentum. For the unpolarized lepton beams, the leptonic tensor $L_{\mu\nu}$ is given by the following: 
\begin{align}
L_{\mu\nu} = l_{1\mu} l_{2\nu} + l_{1\nu} l_{2\mu} - g_{\mu\nu} l_1 \cdot l_2,
\end{align}
with $l_1$ and $l_2$ being the momenta of colliding leptons. The hadronic tensor $W_{\mu\nu}$ contains nonperturbative quantities and is laid out as follows:
\begin{align}
  W^{\mu\nu}
  =&\sum_f |H_f(Q,\mu)^2|^{\mu\nu} \int d^2 \boldsymbol{k}_{A\perp} d^2\boldsymbol{k}_{B\perp}
     \delta^{(2)} ( \boldsymbol{k}_{A\perp}+\boldsymbol{k}_{B\perp}-\boldsymbol{q}_\perp) \nonumber\\
   &\times  \left[D_{1q}^{h_A} (z_A,\boldsymbol{p}_{A\perp};\mu,\zeta_A) 
     D_{1\bar q}^{h_B}(z_B,\boldsymbol{p}_{B\perp};\mu,\zeta_B)+\ldots\right],
\end{align}
where $\boldsymbol{q}_\perp=-\boldsymbol{P}_{A\perp}/z_A$, and $H_f(Q,\mu)$ is the hard scattering factor that can be evaluated in the perturbative QCD. Here, $D_{1q}^{h_A} (z_A,\bm{p}_{A\perp}; \mu,\zeta_A)$ is the TMD FF with $\boldsymbol{p}_{A\perp}$ the transverse momentum of hadron with respect to the fragmenting quark direction, $\mu$ is a renormalization scale, and $\zeta_A$ is a variable to regularize the rapidity divergence. Notice that $\bm{k}_{i,\perp}$ is the relative transverse momentum of the fragmenting parton with respect to the hadron momentum. Therefore, we have $\bm{p}_{i,\perp}$ by $\bm{p}_{i,\perp} = -z_i \bm{k}_{i,\perp}$. Please also notice the difference between $\bm{P}_{A\perp}$ and $\bm{p}_{A\perp}$. The three-dot symbol stands for various spin-dependent terms which are not explicitly shown.

It is more convenient to perform the TMD evolution in the coordinate space than in the momentum space. Therefore, we need the  Fourier transform,
\begin{align}
    D_{1q}^{h} (z,\boldsymbol{p}_{\perp};\mu,\zeta)=\frac{1}{z^2}\int d^2p_{\perp} e^{i \frac{1}{z}\boldsymbol{b}_T\cdot\boldsymbol{p}_\perp} \tilde D_{1q}^h (z,\boldsymbol{b}_T;\mu,\zeta),
\end{align}
to translate the TMD FF into  coordinate space one. The hadronic tensor then becomes the following:
\begin{align}
  W^{\mu\nu}
  =&\sum_f |H_f(Q,\mu)^2|^{\mu\nu} \int \frac{d^2\boldsymbol{b}_T}{(2\pi)^2} e^{-i\boldsymbol{q}_\perp \cdot\boldsymbol{b}_T}
     \left[\tilde D_{1q}^{h_A} (z_A,\boldsymbol{b}_{T};\mu,\zeta_A)  
           \tilde D_{1\bar q}^{h_B}(z_B,\boldsymbol{b}_{T};\mu,\zeta_B)+\ldots\right].
\end{align}

The TMD FF in the coordinate space is defined as the product of transition matrix elements between the vacuum and the hadronic final states. 

Before presenting the final definition for the TMD FF in the coordinate space, we first show the unsubtracted version, which appears in the LO calculation. For the production of hadron A, it reads as follows:
\begin{align}
  &\tilde D_{1q}^{h_A,{\rm unsub}} (z,\boldsymbol{b}_T;y_{p_A}-y_B)
    =\frac{1}{4N_{c}}{\rm Tr}_C{\rm Tr}_D\frac{1}{z} \sum_X \int\frac{dx^-}{2\pi}e^{ik^+x^-}\langle 0|\gamma^+ {\cal L}(\frac{x}{2};+\infty,n_B)\nonumber\\
  &\phantom{XXXXXXXXXXXXXXX}
    \times\psi_q (\frac{x}{2})|h_1,X\rangle\langle h_1,X|\bar\psi_q (-\frac{x}{2}){\cal L}(-\frac{x}{2};+\infty,n_B)^\dagger|0\rangle,
\end{align}
where the position vector $x = (0,x^-, \bm{b}_T)$ contains only minus and transverse components,  $y_{p_A} = \frac{1}{2}\ln \frac{2(p_A^+)^2}{m_A^2}$ is the rapidity of hadron A,  ${\rm Tr}_C$ is a trace in the color space, and ${\rm Tr}_D$ is a trace in the Dirac space. The direction of the Wilson lines in the FF of hadron A is specified by the direction of hadron B which is denoted as $n_B$ and vice versa. Notice that the rapidity parameters $y_A \to +\infty$ and $y_B \to -\infty$ are introduce, so that $n_A= (1, -e^{-2y_A}, \bm{0}_T)$ and $n_B= (-e^{2y_B}, 1, \bm{0}_T)$ are slightly space-like. Please also notice the difference between $y_{p_A}$ and $y_{A}$. The Wilson line starting from the position $x$ is defined as follows:
\begin{align}
    {\cal L}(x;+\infty,n)_{ab}={\cal P}\left\{e^{-ig_0 \int_0^{+\infty} d\lambda n\cdot A_{(0)}^\alpha (x+\lambda n)t^\alpha }\right\}_{ab}.
\end{align}
with $a$ and $b$ being the color indices, and $g_0$ and $A_{(0)}^\alpha$ the bare coupling and the bare \linebreak  gluon field. 

Taking the $y_A \to +\infty$ and $y_B \to -\infty$ limit and absorbing the soft factors into the unsubtracted TMD FF, we arrive at the final definition of the TMD FF:
\begin{align}
  &\tilde D_{1q}^{h_A}(z,\boldsymbol{b}_T;\mu,\zeta_A)= \tilde D_{1q}^{h_A,{\rm unsub}} (z,\boldsymbol{b}_T;y_{p_A}-(-\infty))
  \nonumber \\
  & \phantom{XXXXXXXXXXX} \times
  \sqrt{\frac{\tilde S_{(0)}(\boldsymbol{b}_T;+\infty,y_n)}
  {\tilde S_{(0)}(\boldsymbol{b}_T;+\infty,-\infty)\tilde S_{(0)}(\boldsymbol{b}_T;y_n, -\infty)}} \times Z_D Z_2,
\label{eq:tmd-ff-def}
\end{align}
where $y_n$ is an arbitrary rapidity introduced to separate $\zeta_A \equiv \frac{m_A^2}{z_A^2} e^{2(y_{p_A} - y_n)}$ from $\zeta_B \equiv \frac{m_B^2}{z_B^2} e^{2(y_n - y_{p_B})}$, and $Z_D,Z_2$ are renormalization factors. The bare soft factor $\tilde S_{(0)}$ is defined as the expectation values of Wilson lines on the vacuum, reading as follows:
\begin{align}
  \tilde S_{(0)}(\boldsymbol{b}_T;y_A,y_B)=&
    \frac{1}{N_c}\langle 0|{\cal L}(\frac{\boldsymbol{b}_T}{2};+\infty, n_B)^\dagger_{ca}
                           {\cal L}(\frac{\boldsymbol{b}_T}{2};+\infty, n_A)_{ad}\nonumber \\
         &\times           {\cal L}(-\frac{\boldsymbol{b}_T}{2};+\infty, n_B)_{bc}
                           {\cal L}(-\frac{\boldsymbol{b}_T}{2};+\infty, n_A)^\dagger_{db} |0\rangle.
\end{align}

\subsection{Evolution Equations for TMD FFs}

To regularize the ultraviolet (UV) and rapidity divergences, the energy scale $\mu$ and $\sqrt{\zeta}$ are introduced. As a consequence, the TMD FFs differ at different energy scales. The evolution effects are important for phenomenological studies. The QCD evolution for TMD FFs with respect to $\zeta$ is controlled by the Collins--Soper (CS) equation~\cite{Collins:1981uk,Collins:2011zzd}, which is \linebreak  given as follows:
\begin{align}
  \frac{\partial \ln \tilde D_{1q}^{h}(z,\boldsymbol{b}_T;\mu,\zeta)}{\partial\ln\sqrt{\zeta}}=\tilde K(\boldsymbol{b}_T;\mu),
\end{align}
with $\tilde K(\boldsymbol{b}_T;\mu)$ being the CS evolution kernel. The scale dependence of the evolution kernel is governed by  
\begin{align}
\frac{d\tilde K(\boldsymbol{b}_T;\mu)}{d\ln\mu} &=-\gamma_K(\mu),
\end{align}
where $\gamma_K(\mu)$ is the anomalous dimension. It is given by $\gamma_K (\mu)= \frac{2C_F}{\pi} \alpha_s (\mu)$ with $C_F=4/3$ being the color factor and $\alpha_s$ being the running coupling at the LO accuracy \cite{Aybat:2011zv}.

The $\mu$ dependence of the TMD FF is then given by
\begin{align}
  \frac{d\ln\tilde D_{1q}^{h}(z,\boldsymbol{b}_T;\mu,\zeta)}{\partial\ln\mu}&=\gamma_D(\mu; \frac{\zeta}{\mu^2}) ,
\end{align}
where $\gamma_D$ is another anomalous dimension. At the LO accuracy \cite{Aybat:2011zv}, it is given as follows: $\gamma_D(\mu, \zeta/\mu^2) = \alpha_s (\mu)\frac{C_F}{\pi} (\frac{3}{2} - \ln \frac{\zeta}{\mu^2})$.

The TMD FF defined by Equation~(\ref{eq:tmd-ff-def}) is actually calculable in the colinear factorization approach in the small-$b_T$ regime. However, at the large $b_T$ region, the discrepancy of these two approaches grows in terms of $\Lambda_{\rm QCD} b_T$. This region is usually referred to as the nonperturbative regime since a large coordinate corresponds to a small energy scale. The perturbative treatment of the QCD evolution in this region is no longer reliable. To have a consistent formula, the $b_*$-prescription is usually adopted in phenomenology. By introducing $b_* = |\bm{b}_T|/\sqrt{1+\bm{b}_T^2/b_{\rm max}^2}$ and $\mu_b=2e^{-\gamma_E}/b_*$, we can separate the perturbative part from the nonperturbative part in the QCD evolution. Here, $\gamma_E$ is the Euler constant, and $b_{\rm max}$ is an infrared cutoff which is properly chosen to guarantee that $\mu_b \gg \Lambda_{\rm QCD}$. Employing the $b_*$ prescription, the QCD evolution is always performed in the realm of the perturbative QCD. Therefore, this approach underestimates the contribution from the nonperturbative regime. This part of the contribution can be reintegrated into the final prescription by the introduction of a nonperturbative factor. 

Ultimately, we arrive at \cite{Aybat:2011zv}
\begin{align}
 D_{1q}^{h} (z, b_T; \mu, \zeta) &= D_{1q}^{h} (z, b_T^*; \mu_0=\mu_b, \zeta_0=\mu_b^2) \nonumber\\
 &\times \exp\bigg\{
 \ln\frac{\sqrt{\zeta}}{\mu_b}\tilde K(b_*;\mu_b)+\int_{\mu_b}^\mu \frac{d\mu^\prime}{\mu^\prime}\gamma_D(\mu^\prime; \frac{\zeta}{\mu'^2})\bigg\} \nonumber \\
 &\times \exp\bigg\{-S_{\rm np} (z, |\bm{b}_T|, \zeta)\bigg\},
\end{align}
where the last line is the nonperturbative function that returns the nonperturbative effect that has been deliberately removed from the QCD evolution in the $b_*$-prescription. There is no theoretical approach that can evaluate this nonperturbative function other than the one that extracts it from experimental data \cite{Davies:1984sp,Landry:2002ix,Guzzi:2013aja,Sun:2014wpa,Ladinsky:1993zn,Sun:2013hua,Aidala:2014hva,Echevarria:2014xaa,Kang:2011mr,Collins:2014jpa}. Notice that ~\cite{Qiu:2000hf} present a different method to address the nonperturbative physics. Here, $D_{1q}^{h} (z, b_T^*; \mu_0=\mu_b, \zeta_0=\mu_b^2)$ is the FF at the initial scale. In the phenomenology, it is usually chosen to coincide with the colinear FF $D_{1q}^{h} (z, \mu_f)$ with the factorization scale specified by $\mu_f=\mu_b$.

Similar to the PDF case, QCD evolution tends to broaden the $k_T$ distribution width at higher energy scales. Both unpolarized and spin-dependent FFs show such a behavior~\cite{Kang:2015msa}.

\subsection{TMD Factorization at the Higher Twist}

\textls[-15]{In a semi-inclusive process, normally we can find two energy scales: the typical transverse momentum $q_\perp$ and the hardest energy scale $Q$. In the region of $Q \gg q_\perp \gg \Lambda_{\rm QCD}$, the TMD factorization framework at the leading twist usually works very well. When $q_\perp \sim Q \gg \Lambda_{\rm QCD}$, we should fall back to the colinear factorization. However, in between, there is still a large phase space where $q_\perp$ is smaller than $Q$ but not much smaller. This is the kinematic region where both the TMD factorization and the colinear factorization can approximately apply. However, the prediction from the TMD factorization deviates from the experimental measurements when $q_\perp/Q$ becomes not very small, calling for the inclusion of higher twist corrections. The higher twist corrections are also usually referred to as power corrections since they provide contributions in terms of $(q_\perp/Q)^n$. In addition, twist-3 contributions usually introduce new asymmetries that do not appear at the leading twist level. A comprehensive study on the higher twist contributions is thus vital in phenomenology. Contributions from higher twist TMD PDFs and FFs were studied some time ago~\cite{Mulders:1995dh}, but few advances have been made in the systematic derivations of the TMD factorization formula at the higher twist gain. Various theoretical methods have been applied to derive the TMD factorization scheme at the twist-3 level, such as the TMD operator expansion technique~\cite{Vladimirov:2021hdn, Rodini:2022wki, Rodini:2022wic}, the soft-colinear effective theory approach~\cite{Ebert:2021jhy}, factorization from functional integral~\cite{Balitsky:2017flc, Balitsky:2017gis, Balitsky:2020jzt, Balitsky:2021fer}, and a very recent work from~\cite{Gamberg:2022lju}, etc. The TMD factorization at the higher twist level is far from being completed, which requires further theoretical efforts.}

\section{Spin-Dependent TMD FFs}
\label{sec:spin}

In semi-inclusive reactions, the experimental observables are usually different azimuthal asymmetries. In the kinematic region of TMD factorization, they are directly linked to TMD PDFs or FFs. The transverse momenta of partons and hadrons are often entangled with their polarizations. As a consequence, there are abundant polarization-dependent azimuthal asymmetries that can be measured in the experiment. This is particularly true for the transverse polarization. It is thought to provide only subleading power contributions compared to the longitudinal polarization at high energy; however, it often generates leading power contributions when correlated with the transverse momenta. In this section, we summarize the definition of the spin-dependent TMD FFs for hadrons with different spins. The following discussion only applies at the LO level since the TMD factorization for higher twist contributions is still far from being concluded. Therefore, we remove the scale dependence from TMD FFs. 

\subsection{The Intuitive Definition of TMD FFs}

FFs represent the momentum distribution of a hadron inside of a hadronic jet produced by the fragmenting high-energy parton. We use ${\cal D}^{q\to h}(k;p)$ to denote the probability density of producing a hadron $h$ with momentum $p$ from a quark with momentum $k$. 

In the high-energy limit, we can safely neglect the quark and hadron mass. Therefore, we have $k^2 = p^2 =0$. In the naive parton model picture, the hadrons move colinearly with the parent quark. We thus have $p=zk$ where $p$ is the hadron momentum, $k$ is the quark momentum, and $z$ is the momentum fraction. In this case, the FF is only a scalar function of $z$. We have
\vspace{-6pt}
\begin{align}
{\cal D}^{q\to h}(k;p) = D_1^{q\to h}(z),
\end{align}
where $D_1^{q\to h}(z)$ is simply the unpolarized FF. 

With the spin degree of freedom being taken into account, the FFs will also depend on additional parameters which characterize the polarization of the final state hadron or the fragmenting quark. For example, for the production of spin-1/2 hadrons, we need to introduce $\lambda_q$ and $\lambda$ to describe the helicities and introduce $\vec s_{Tq}$ and $\vec S_{T}$ to describe the transverse polarizations of the quark and the hadron. With more available parameters, we can construct two additional scalar structures, $\lambda_q \lambda_h$ and $\vec s_{Tq} \cdot \vec S_{T}$, according to the parity conservation. Therefore, the complete decomposition of the FF is given by
\begin{align}
{\cal D}^{q\to h}(k,S_q;p,S) = D_1^{q\to h}(z) + \lambda_q \lambda G_{1L}^{q\to h}(z) + \vec s_{Tq} \cdot \vec S_T H_{1T}^{q\to h}(z), \label{eq:1D-FF}
\end{align}
where $G_{1L}(z)$ and $H_{1T}(z)$ are the longitudinal and transverse spin transfers from the quark to the hadron, respectively. The physical interpretations of these probability densities coincide with those of the leading twist FFs in the colinear factorization approach. 

In some cases, the transverse momentum of the final state hadron with respect to the quark momentum becomes relevant to the observable of interest. The interplay between the transverse momentum $p_\perp$ and the polarization parameters induces considerably intriguing phenomena. Again, we use the spin-1/2 hadron production as an example. From the parton model, we obtain the following eight TMD probability densities:
\begin{align}
{\cal D}(k,S_q;p,S) =& D_1(z,p_\perp) + \lambda_q \lambda G_{1L}(z,p_\perp) + \vec s_{Tq} \cdot \vec S_T H_{1T}(z,p_\perp) \nonumber\\
+& \frac{1}{M} \vec S_T \cdot (\hat {\vec k} \times \vec p_\perp) D_{1T}^\perp(z,p_\perp) + \frac{1}{M} \lambda_q (\vec S_T \cdot \vec p_\perp) G_{1T}^\perp(z,p_\perp) \nonumber\\
+& \frac{1}{M} \vec s_{Tq} \cdot (\hat {\vec k} \times \vec p_\perp) H_1^\perp(z,p_\perp) + \frac{1}{M} \lambda (\vec s_{Tq} \cdot \vec p_\perp) H_{1L}^\perp(z,p_\perp) \nonumber \\
+& \frac{1}{M^2} (\vec s_{Tq} \cdot \vec p_\perp)(\vec S_T \cdot \vec p_\perp) H_{1T}^\perp(z,p_\perp). \label{eq:3D-FF}
\end{align}

Here, we have dropped the $q\to h$ superscript for simplicity. These TMD FFs correspond to the eight leading twist TMD FFs defined in the TMD factorization approach. Among them, we notice in particular the famous Collins function $H_1^\perp$~\cite{Collins:1992kk} and the Sivers-type FF $D_{1T}^\perp$~\cite{Sivers:1989cc,Sivers:1990fh}. They are usually referred to as the naive-T-odd FFs. In neglecting the interaction among the final state hadrons and the gauge link (which will be explained below), the time-reversal invariance demands that these two functions disappear. However, the time-reversal operation converts the ``out'' state to the ``in'' state. The interaction among hadrons suggests that one cannot find a simple relation between the ``in'' and ``out'' states any longer. Therefore, the time-reversal invariance actually poses no constraints on FFs. This feature can be fully appreciated in the context of parton correlators in the next subsection. Furthermore, we use $H$ to denote FFs accompanied with the transverse polarization of the fragmenting quark $\vec s_{Tq}$. They are chiral-odd FFs. The reason for this will also be explained later. 

\subsection{The Definition of TMD FFs from the Parton Correlators}

In the language of quantum field theory, the quark FFs are defined via the decomposition of parton correlators, such as the quark--quark correlator and the quark--gluon correlator. Usually, we need to define the gauge-invariant quark--quark correlators in the very beginning. From \cite{Collins:1981uw, Mulders:1995dh,Boer:1997mf,Bacchetta:2006tn,Wei:2014pma}, we have the following:
\begin{align}
\hat\Xi_{ij}^{(0)}(k;p,S) =\frac{1}{2\pi} & \sum_X \int  d^4\xi e^{-ik \xi} \langle 0| \mathcal{L}^\dag (0;\infty) \psi_i(0) |p,S;X\rangle  \langle p,S;X|\bar\psi_j(\xi) \mathcal{L}(\xi;\infty) |0\rangle, \label{eq:qq-correlator}
\end{align}
where $\xi$ is the coordinate of the quark field, $k$ and $p$ denote the 4-momenta of the fragmenting quark and the produced hadron, respectively; $S$ denotes the hadron spin; and $\mathcal{L}(\xi;\infty)$ is the gauge link that ensures the gauge invariance of the definition of the correlator. We use $i$ and $j$ to represent one component of the corresponding spinor. Therefore, $\hat\Xi_{ij}^{(0)}(k;p,S)$ is actually one element in a $4 \times 4$ matrix which is denoted by $\hat\Xi^{(0)}(k;p,S)$. 

As for the TMD FFs, we can integrate the above master correlator over the $k^-$ component and obtain the following TMD quark--quark correlator:
\begin{align}
\hat\Xi_{ij}^{(0)}(z,k_{\perp};p,S) =
\sum_X &\int \frac{p^+d\xi^-}{2\pi} d^2{\xi}_\perp e^{-i(p^+\xi^-/z - \vec{k}_{\perp} \cdot \vec{\xi}_\perp)} \nonumber\\
&\times \langle 0| \mathcal{L}^\dag (0;\infty) \psi_i(0) |p,S;X\rangle \langle p,S;X|\bar\psi_j(\xi) \mathcal{L}(\xi;\infty) |0\rangle, \label{eq:Xi0}
\end{align}
where $z=p^+/k^+$ is the longitudinal momentum fraction of the hadron, and $k_\perp$ is the transverse momentum of the fragmenting quark with respect to the hadron momentum. Unlike the discussion in the previous sections, it is more convenient to express the parton correlators as a function of $k_\perp$ instead of $p_\perp$. Nonetheless, since we have the approximation $\bm{k}_\perp = - \bm{p}_\perp/z$, these two methods are equivalent. 

Although the TMD quark--quark correlator is a nonperturbative object, we can still discuss some general features from the definition. For instance, it possesses hermiticity, parity invariance, and charge-conjugation symmetry. As will be shown below, these properties will constrain the structures of the correlator. However, unlike the case for PDFs, the time-reversal invariance does not mean much for FFs. 

Furthermore, the quark--quark correlator is a $4\times 4$ matrix in the Dirac space. Therefore, it can always be decomposed in terms of 16 $\Gamma$-matrices, i.e.,
\begin{align}
\hat\Xi^{(0)}(z,k_{\perp};p,S) &=  \Xi^{(0)}(z,k_{\perp};p,S) + i\gamma_5 \tilde\Xi^{(0)}(z,k_{\perp};p,S) + \gamma^\alpha \Xi_\alpha^{(0)}(z,k_{\perp};p,S) \nonumber\\
&+ \gamma_5\gamma^\alpha \tilde\Xi_\alpha^{(0)}(z,k_{\perp};p,S) + i\sigma^{\alpha\beta}\gamma_5 \Xi_{\alpha\beta}^{(0)}(z,k_{\perp};p,S). \label{eq:XiExpansion}
\end{align}
The coefficient functions $\Xi^{(0)}$, $\tilde\Xi^{(0)}$,  $\Xi_\alpha^{(0)}$, $\tilde\Xi_\alpha^{(0)}$ and $\Xi_{\alpha\beta}^{(0)}$ are given by the trace of the corresponding $\Gamma$-matrix with the correlator. These coefficient functions can further be decomposed into the products of scalar functions with basic Lorentz covariants according to their Lorentz transformation properties. The basic Lorentz covariants are constructed in terms of the available kinematic variables used in the reaction process. The scalar functions are the corresponding TMD FFs. We will present the detailed decomposition in the following subsections. Notice that the TMD quark--quark correlator given by Equation (\ref{eq:Xi0}) satisfies the constraints of hermiticity and parity conservation. This will limit the allowed Lorentz structures of the parton correlator.

Higher twist TMD FFs also receive contributions from quark--gluon \linebreak correlators \cite{Mulders:1995dh,Boer:1997mf,Bacchetta:2006tn,Wei:2014pma} in addition to the quark--quark correlator mentioned above. For example, the complete decomposition of twist-3 TMD FFs also involves contributions from the following correlator:
\vspace{-6pt}
\begin{adjustwidth}{-\extralength}{0cm}
\begin{align}
\hat\Xi_{\rho,ij}^{(1)}(k;p,S)=&  \sum_X \int  \frac{d^4\xi}{2\pi} e^{-ik\cdot \xi}  \langle 0| \mathcal{L}^\dag (0;\infty) D_\rho(0)\psi_i(0) |p,S;X\rangle \langle p,S;X|\bar\psi_j(\xi) \mathcal{L}(\xi;\infty) |0\rangle,  \label{eq:qgq-correlator}
\end{align}
\end{adjustwidth}
where $D_\rho(y)\equiv -i\partial_\rho+gA_\rho(y)$ and $A_\rho(y)$ denote the gluon field. However, the twist-3 TMD FFs defined via these quark--gluon correlators are not independent from those defined via the quark--quark correlator~\cite{Wei:2014pma,Chen:2016moq}. They are related to each other by a set of equations derived using the QCD equation of motion $\gamma \cdot D(y)\psi(y)=0$. Therefore, we will only show the explicit decomposition of the TMD FFs from the quark--quark correlator in the \linebreak following subsections.

\subsection{The Spin Dependence}

With the spin degree of freedom being taken into account, the basic Lorentz covariants in the decompositions of the coefficient functions in Equation~(\ref{eq:XiExpansion}) depend on not only momenta but also parameters describing the hadron polarization. The hadron polarization is defined in the rest frame of the hadron and is described by the spin density matrix.

For spin-1/2 hadrons, the spin density matrix is given by
\begin{align}
\rho = \frac{1}{2}\left( 1 + \vec S \cdot \vec \sigma \right),
\end{align}
where $\vec \sigma$ is the Pauli matrix, and $\vec S$ is the polarization vector in the rest frame of the hadron. The covariant form of the polarization vector reads as follows:
\begin{align}
S^\mu = \lambda \frac{p^+}{M} \bar n^\mu + S_T^\mu - \lambda \frac{M}{2p^+} n^\mu. \label{eq:PolVector}
\end{align}

Here, $M$ is the hadron mass, $\lambda$ is the helicity, and $S_T^\mu$ is the transverse polarization vector of the hadron. We have employed $\bar n^\mu$ to represent the light-cone plus direction and $n^\mu$ to denote the minus direction. For spin-1/2 hadrons, an additional pseudo-scalar $\lambda$ and an axial-vector $S_T^\mu$ are at our disposal for constructing the basic Lorentz tensors.

For spin-1 hadrons, such as the vector mesons, the polarization is described by a $3\times 3$ density matrix, which is usually given as \cite{Bacchetta:2000jk}
\begin{align}
\rho = \frac{1}{3} (\mathbf{1} + \frac{3}{2}S^i \Sigma^i + 3 T^{ij} \Sigma^{ij}). \label{eq:spin1rho}
\end{align}

Here, $\Sigma^i$ is the spin operator of spin-$1$ particle. The rank-2 tensor polarization basis $\Sigma^{ij}$ is defined by
\begin{align}
\Sigma^{ij} \equiv \frac{1}{2} (\Sigma^i\Sigma^j + \Sigma^j \Sigma^i) - \frac{2}{3} \mathbf{1} \delta^{ij}.
\label{eq:Sigmaij1}
\end{align}
where the second term subtracts the diagonal elements from the product in the first term to give the relation
\begin{align}
\Sigma^{x x}+\Sigma^{y y}+\Sigma^{z z}&=0.
\label{eq:Sigmaxxyyzz}
\end{align}

This can be easily seen from the square of the spin-1 operator, i.e., $\Sigma^2 \equiv \Sigma^x \Sigma^x + \Sigma^y \Sigma^y + \Sigma^z \Sigma^z = s(s+1){\bm 1}$ with $s=1$ for spin-1. From Equation~(\ref{eq:spin1rho}), we find that a polarization tensor $T$ is required to fully describe the polarization of a vector meson besides the polarization vector $S$. The polarization vector $S$ is similar to that of spin-1/2 hadrons. It takes the same covariant form as laid out in Equation~(\ref{eq:PolVector}). The polarization tensor $T^{ij}={\rm Tr}(\rho \Sigma^{ij})$ has five independent components that consist of a Lorentz scalar $S_{LL}$, a Lorentz vector $S_{LT}^\mu = (0, S_{LT}^x, S_{LT}^y,0)$ and a Lorentz tensor $S_{TT}^{\mu\nu}$ that has two nonzero independent components ($S_{TT}^{xx} = -S_{TT}^{yy}$ and $S_{TT}^{xy} = S_{TT}^{yx}$). It is parameterized as follows:
\begin{align}
\bm{T}= \frac{1}{2}
\left(
\begin{array}{ccc}
-\frac{2}{3}S_{LL} + S_{TT}^{xx} & S_{TT}^{xy} & S_{LT}^x  \\
S_{TT}^{xy}  & -\frac{2}{3} S_{LL} - S_{TT}^{xx} & S_{LT}^{y} \\
S_{LT}^x & S_{LT}^{y} & \frac{4}{3} S_{LL}
\end{array}
\right).
\label{spintensor}
\end{align}

The Lorentz covariant form for the polarization tensor is expressed as~\cite{Bacchetta:2000jk}
\begin{align}
T^{\mu\nu} = \frac{1}{2} & \Bigl[ \frac{4}{3}S_{LL}\Bigl( \frac{p^+}{M} \Bigr)^2 \bar n^\mu \bar n^\nu + \frac{p^+}{M} n^{\{\mu}S_{LT}^{\nu\}} 
- \frac{2}{3}S_{LL}(\bar n^{\{\mu}n^{\nu\}} - g_T^{\mu\nu}) \nonumber\\
&+ S_{TT}^{\mu\nu} 
- \frac{M}{2p^+} \bar n^{\{\mu}S_{LT}^{\nu\}} + \frac{1}{3}S_{LL}\Bigl( \frac{M}{p^+} \Bigr)^2 n^\mu n^\nu \Bigr],
\end{align}
where we have used the shorthand notation $A^{\{\mu}B^{\nu\}} \equiv A^\mu B^\nu + A^\nu B^\mu$.

For spin-$3/2$ hadrons, such as the decuplet baryons, the polarization is described by a $4\times 4$ density matrix which is given by~\cite{Song:1967, Zhao:2022lbw}
\begin{align}
\rho=\frac{1}{4}\left(\bm{1}+\frac{4}{5} S^{i} \Sigma^{i}+\frac{2}{3} T^{i j} \Sigma^{i j}+\frac{8}{9}R^{i j k}\Sigma^{i j k}\right). \label{eq:density-3-2}
\end{align}

Here, $\Sigma^i$ is the spin operator of the spin-$3/2$ particle, and $S^i$ is the corresponding polarization vector.
Similar with that for spin-$1$ case, $(\Sigma^{ij})$ is the polarization tensor basis which has five independent components. It can be constructed from $\Sigma^i$ and is given by
\begin{align}
\Sigma^{i j}&=\frac{1}{2}\left(\Sigma^{i} \Sigma^{j}+\Sigma^{j} \Sigma^{i}\right)-\frac{5}{4} \delta^{i j}{\bm 1},
\label{eq:Sigmaij32}
\end{align}

Notice that the square of the spin-3/2 operator is given by $\sum_i (\Sigma^i)^2 = \frac{3}{2}(\frac{3}{2}+1) {\bm 1} = \frac{15}{4} {\bm 1}$. The rank-2 tensor polarization basis for spin-3/2, $\Sigma^{ij}$, is also chosen to be traceless as laid out by Equation~(\ref{eq:Sigmaxxyyzz}). Therefore, the second term in Equation~(\ref{eq:Sigmaij32}) is different from that in Equation~(\ref{eq:Sigmaij1}) for spin-1 hadrons. The corresponding polarization tensor $T^{ij}$ also has five independent components which are the same as those for spin-$1$ hadrons. The rank-3 tensor polarization basis $\Sigma^{ijk}$ is unique for spin-$3/2$ hadrons. It has seven independent components which can be constructed as follows:
\begin{align}
    \Sigma^{i j k}&=\frac{1}{6} \Sigma^{\{i} \Sigma^{j}\Sigma^{k\}}
    -\frac{41}{60}  \left(\delta^{ij}\Sigma^k + \delta^{jk}\Sigma^i + \delta^{ki}\Sigma^j\right) \notag \\
    &= \frac{1}{3}\left(\Sigma^{ij}\Sigma^{k} + \Sigma^{jk}\Sigma^{i} + \Sigma^{ki}\Sigma^{j} \right) 
    - \frac{4}{15}\left(\delta^{ij}\Sigma^k + \delta^{jk}\Sigma^i + \delta^{ki}\Sigma^j \right),
    \label{e.Sigmaijk}
\end{align}
where the symbol $\{\cdots\}$ stands for the sum of all possible permutations.
The corresponding rank-3 spin tensor $R^{ijk}$ is defined as follows:
\begin{equation}
R^{i j k}=\frac{1}{4}
\left[
\begin{aligned}
&\left(
\begin{array}{ccc}
-3S_{LLT}^x+S_{TTT}^{xxx} & -S_{LLT}^y+S_{TTT}^{yxx} & -2S_{LLL}+S_{LTT}^{xx}\\
 -S_{LLT}^y+S_{TTT}^{yxx} & -S_{LLT}^x-S_{TTT}^{xxx} & S_{LTT}^{xy}\\
 -2S_{LLL}+S_{LTT}^{xx} &S_{LTT}^{xy} & 4S_{LLT}^{x}
\end{array}
\right)\\
&\left(
\begin{array}{ccc}
-S_{LLT}^y+S_{TTT}^{yxx} & -S_{LLT}^x-S_{TTT}^{xxx} & S_{LTT}^{xy}\\
-S_{LLT}^x-S_{TTT}^{xxx} & -3S_{LLT}^y-S_{TTT}^{yxx} & -2S_{LLL}-S_{LTT}^{xx}\\
S_{LTT}^{xy} & -2S_{LLL}-S_{LTT}^{xx} & 4S_{LLT}^{y}
\end{array}
\right)\\
&\left(
\begin{array}{ccc}
 -2S_{LLL}+S_{LTT}^{xx} & S_{LTT}^{xy} & 4S_{LLT}^{x}\\
 S_{LTT}^{xy} & -2S_{LLL}-S_{LTT}^{xx} & 4S_{LLT}^{y}\\
 4S_{LLT}^{x} & 4S_{LLT}^{y} & 4S_{LLL}
\end{array}
\right)
\end{aligned}
\right],
\label{eq:tensor2}
\end{equation}
Meanwhile, the Lorentz covariant form is given as follows:
\vspace{-6pt}
\begin{adjustwidth}{-\extralength}{0cm}
\begin{align}
R^{\mu \nu \rho}&=
\frac{1}{4}\Bigg\{S_{LLL}\Bigg[\frac{1}{2}\left(\frac{M}{P \cdot \bar{n}}\right)^3 \bar{n}^\mu\bar{n}^\nu\bar{n}^\rho
-\frac{1}{2}\left(\frac{M}{P \cdot \bar{n}}\right)\left(\bar{n}^{\{\mu} \bar{n}^\nu n^{\rho\}}
-\bar{n}^{\{\mu}g_{T}^{\nu\rho\}} \right)       \nonumber\\
&\quad +\left(\frac{P \cdot \bar{n}}{M}\right)\left(\bar{n}^{\{\mu} n^\nu n^{\rho\}}-n^{\{\mu}g_{T}^{\nu\rho\}} \right)-4\left(\frac{P \cdot \bar{n}}{M}\right)^3 n^{\mu} n^\nu n^{\rho}\Bigg]     \nonumber\\
&\quad +\frac{1}{2}\left(\frac{M}{P \cdot \bar{n}}\right)^2 \bar{n}^{\{\mu} \bar{n}^\nu S_{LLT}^{\rho\}}+2\left(\frac{P \cdot \bar{n}}{M}\right)^2 n^{\{\mu} n^\nu S_{LLT}^{\rho\}}-2\bar{n}^{\{\mu} n^\nu S_{LLT}^{\rho\}}+\frac{1}{2}S_{LLT}^{\{\mu}g_{T}^{\nu\rho\}}        \nonumber\\
&\quad +\frac{1}{4}\left(\frac{M}{P \cdot \bar{n}}\right) \bar{n}^{\{\mu}  S_{LTT}^{\nu\rho\}}-\frac{1}{2}\left(\frac{P \cdot \bar{n}}{M}\right) n^{\{\mu}  S_{LTT}^{\nu\rho\}}+S_{TTT}^{\mu\nu\rho}\Bigg\}.
\label{eq:spin_R}
\end{align}
\end{adjustwidth}

\subsection{Decomposition Result for Spin-Dependent TMD FFs}

The results for TMD FFs of spin-1 hadrons defined via quark--quark correlator exist up to twist-4 level in the literature~\cite{Chen:2016moq}. The leading twist TMD FFs for spin-3/2 hadrons have also been presented in~\cite{Zhao:2022lbw}. In this section, we summarize the general decomposition of the quark--quark correlator in terms of TMD FFs for the unpolarized part, polarization-vector-dependent part, rank-2-polarization-tensor-dependent part, and rank-3-polarization-tensor-dependent parts. To describe the production of pseudoscalar mesons, we only need the unpolarized part. To describe the production of baryons, we need to combine the unpolarized and the polarization-vector-dependent parts. The description of the spin-$3/2$ hadron production requires all four parts. However, it should be noted that different conventions are employed in different works. 

The notation system for TMD FFs in this review are laid out here. We use $D$, $G$, and $H$ to denote FFs of unpolarized, longitudinally polarized, and transversely polarized quarks, respectively. They are obtained from the decomposition of the $\gamma_\mu$, $\gamma_5\gamma_\mu$ and $\gamma_5 \sigma_{\mu\nu}$ terms of the quark--quark correlator. Those FFs defined from the decomposition of the $\bm{1}$ and $\gamma_5$ terms are denoted as $E$. We use the numbers $1$ and $3$ in the subscripts to denote the leading twist and twist-4 FFs, respectively. Other FFs without numbers in the subscripts are at the twist-3 level. The polarization of the produced hadron will be specified in the subscripts, where $L$ and $T$ represent longitudinal and transverse polarizations, and $LL$, $LT$, and $TT$ stand for the rank-2-tensor polarizations. The symbol $\perp$ in the superscript implies that the corresponding basic Lorentz structure depends on the transverse momentum $k_\perp$.

The decomposition for the unpolarized part is given by the following:
\begin{align}
& z\Xi^{U(0)}(z,k_{\perp};p) = ME(z,k_{\perp}), \label{eq:XiUS}\\
& z\tilde\Xi^{U(0)}(z,k_{\perp};p) =0, \label{eq:XiPS}\\
& z\Xi_\alpha^{U(0)}(z,k_{\perp};p) = p^+ \bar n_\alpha D_1(z,k_{\perp})+ k_{\perp\alpha} D^\perp(z,k_{\perp}) + \frac{M^2}{p^+}n_\alpha D_3(z,k_{\perp}),\label{eq:XiUV}\\
 & z\tilde\Xi_\alpha^{U(0)}(z,k_{\perp};p) = -\tilde k_{\perp\alpha} G^\perp(z,k_{\perp}), \label{eq:XiUAV}\\
& z\Xi_{\rho\alpha}^{U(0)}(z,k_{\perp} ;p) = -\frac{p^+}{M} \bar n_{[\rho}\tilde k_{\perp\alpha]} H_1^\perp(z,k_{\perp}) + M\varepsilon_{\perp\rho\alpha} H(z,k_{\perp}) 
- \frac{M}{p^+} n_{[\rho}\tilde k_{\perp\alpha]} H_3^\perp(z,k_{\perp}).  \label{eq:XiUT}
\end{align}

Here, $\tilde k_{\perp\alpha} \equiv \varepsilon_{\perp\mu\alpha}k_{\perp}^\mu$ denotes the transverse vector orthogonal to $k_{\perp\alpha}$, with $\epsilon_{\perp\mu\nu}$ being defined as $\varepsilon_{\perp\mu\nu} \equiv \varepsilon_{\mu\nu\alpha\beta} \bar n^\alpha n^\beta$. There are eight TMD FFs for the unpolarized part. Among them, the number density $D_1$ and the Collins function $H_1^\perp$ are at the leading twist. They both have twist-4 companions i.e., $D_3$ and $H_3^\perp$, respectively. The other four are twist-3 FFs. The TMD FFs $D_{1T}^\perp$,$G^\perp$, $H_1^\perp$, $H$, and $H_3^\perp$ are usually referred to as the naive T-odd FFs. The reader may have already discerned that the T-odd FFs are always associated with the Levi-Civita tensor, $\varepsilon_{\mu\nu\alpha\beta}$. It should be noted that T-odd PDFs can only survive thanks to the gauge link. However, for the FFs, the final state interactions between the produced hadrons in the hadronization process can also contribute to the T-oddness. This difference has a more important impact on the polarization-vector-dependent T-odd PDFs and FFs, which are discussed below. 

The decomposition for the vector polarized part is given by the following:

\vspace{-6pt}
\begin{adjustwidth}{-\extralength}{0cm}
\begin{align}
z\Xi^{V(0)}(z,k_{\perp};p,S) &= (\tilde k_{\perp} \cdot S_T) E_T^\perp(z,k_{\perp}), \label{eq:XiVS}\\
z\tilde \Xi^{V(0)}(z,k_{\perp};p,S) &= M \Bigl[ \lambda E_L(z,k_{\perp}) + \frac{k_{\perp} \cdot S_T}{M} E^{\prime\perp}_T(z,k_{\perp}) \Bigr],  \label{eq:XiVPS}\\
z\Xi_\alpha^{V(0)}(z,k_{\perp};p,S) &= p^+ \bar n_\alpha \frac{\tilde k_{\perp} \cdot S_T}{M} D_{1T}^\perp(z,k_{\perp}) 
 - M\tilde S_{T\alpha} D_T(z,k_{\perp}) \nonumber\\
& \hspace{-0.5cm} - \tilde k_{\perp\alpha} \Bigl[ \lambda D_L^\perp(z,k_{\perp}) + \frac{k_{\perp} \cdot S_T}{M}D_T^{\perp}(z,k_{\perp}) \Bigr] + \frac{M}{p^+}n_\alpha (\tilde k_{\perp} \cdot S_T) D_{3T}^\perp(z,k_{\perp}),  \label{eq:XiVV}\\
z\tilde\Xi_\alpha^{V(0)}(z,k_{\perp};p,S) &= p^+ \bar n_\alpha \Bigl[ \lambda G_{1L}(z,k_{\perp}) + \frac{k_{\perp} \cdot S_T}{M} G_{1T}^\perp(z,k_{\perp}) \Bigr] \nonumber\\
& \hspace{-0.5cm} - MS_{T\alpha} G_T(z,k_{\perp}) - k_{\perp\alpha} \Bigl[ \lambda G_{L}^\perp(z,k_{\perp}) + \frac{k_{\perp} \cdot S_T}{M} G_{T}^{\perp}(z,k_{\perp}) \Bigr] \nonumber\\ 
& \hspace{-0.5cm} + \frac{M^2}{p^+}n_\alpha \Bigl[ \lambda G_{3L}(z,k_{\perp}) + \frac{k_{\perp} \cdot S_T}{M} G_{3T}^\perp(z,k_{\perp}) \Bigr],  \label{eq:XiVAV}\\
z\Xi_{\rho\alpha}^{V(0)}(z,k_{\perp};p,S) &= p^+ \bar n_{[\rho}S_{T\alpha]} H_{1T}(z,k_{\perp}) + \frac{p^+}{M} \bar n_{[\rho}k_{\perp\alpha]} \Bigl[ \lambda H_{1L}^\perp(z,k_{\perp}) 
+ \frac{k_{\perp} \cdot S_T}{M} H_{1T}^\perp(z,k_{\perp}) \Bigr] \nonumber\\
& \hspace{-1.2cm} + k_{\perp[\rho}S_{T\alpha]} H_T^\perp(z,k_{\perp}) + M \bar n_{[\rho}n_{\alpha]} \Bigl[ \lambda H_L(z,k_{\perp}) 
+ \frac{k_{\perp} \cdot S_T}{M} H_T^{\prime\perp}(z,k_{\perp}) \Bigr] \nonumber\\
& \hspace{-1.2cm} + \frac{M^2}{p^+} n_{[\rho}S_{T\alpha]} H_{3T}(z,k_{\perp}) 
+ \frac{M}{p^+} n_{[\rho}k_{\perp\alpha]} \Bigl[ \lambda H_{3L}^\perp(z,k_{\perp}) + \frac{k_{\perp} \cdot S_T}{M} H_{3T}^\perp(z,k_{\perp}) \Bigr].  \label{eq:XiVT}
\end{align}
\end{adjustwidth}

There are in total 24 polarization-vector-dependent TMD FFs. Of these, 6 contribute at the leading twist, 12 at twist-3, and remaining 6 at twist-4.
Among the six leading twist FFs, $G_{1L}$ is the longitudinal spin transfer, $H_{1T}$ and $H_{1T}^\perp$ are transverse spin transfers, $G_{1T}^\perp$ is the longitudinal to transverse spin transfer, $H_{1L}^\perp$ is the transverse to longitudinal spin transfer, and $D_{1T}^\perp$ induces the transverse polarization of hadrons in the fragmentation of an unpolarized quark. We note in particular that the $D_{1T}^\perp$ FF resembles the Sivers function in PDFs~\cite{Sivers:1989cc}. It is responsible for the hadron transverse polarization along the normal direction of the production plane in high-energy collisions. It is also a naive T-odd FF. However, as mentioned above, the T-oddness has little meaning in the context of hadronization. The T-odd PDFs arise solely from the gauge link. Therefore, it has been proven theoretically that there is a sign-flip between the Sivers functions in SIDIS and Drell-Yan~\cite{Collins:2002kn,Ji:2002aa,Belitsky:2002sm}. However, the T-oddness of FFs can also be generated from the interaction among final state hadrons. Therefore, there is no such similar relation for the $D_{1T}^\perp$ FF between different processes. Besides $D_{1T}^\perp$, there are seven other T-odd FFs, namely,  $E_T^\perp$, $E_L$, $E_T^{\prime\perp}$, $D_L^\perp$, $D_T$, $D_T^{\perp}$, and $D_{3T}^\perp$. The rest are T-even. All of T-odd FFs are accompanied by the Levi-Civita tensor except for $E_L$ and $E_T^{\prime\perp}$.

The decomposition for the rank-2-polarization-tensor-dependent part is given as follows:
\vspace{-6pt}
\begin{adjustwidth}{-\extralength}{0cm}
\begin{align}
& z\Xi^{T(0)}(z,k_{\perp};p,S) =M \Bigl[ S_{LL}E_{LL}(z,k_{\perp}) 
+ \frac{k_{\perp} \cdot S_{LT}}{M} E_{LT}^\perp(z,k_{\perp}) 
+ \frac{S_{TT}^{kk}}{M^2} E_{TT}^{\perp}(z,k_{\perp})\Bigr],  \label{eq:XiTS} \\
& z\tilde \Xi^{T(0)}(z,k_{\perp};p,S) =M \Bigl[ \frac{\tilde k_{\perp} \cdot S_{LT}}{M} E_{LT}^{\prime\perp} (z,k_{\perp}) 
+ \frac{S_{TT}^{\tilde k k}}{M^2} E_{TT}^{\prime\perp}(z,k_{\perp}) \Bigr],  \label{eq:XiTPS}\\
& z\Xi_\alpha^{T(0)}(z,k_{\perp};p,S) = p^+ \bar n_\alpha \Bigl[ S_{LL} D_{1LL}(z,k_{\perp}) 
+ \frac{k_{\perp} \cdot S_{LT}}{M}D_{1LT}^\perp (z,k_{\perp}) + \frac{S_{TT}^{kk}}{M^2} D_{1TT}^\perp(z,k_{\perp}) \Bigr] \nonumber\\
 & \hspace{2cm} + M S_{LT\alpha} D_{LT}(z,k_{\perp}) +S_{TT\alpha}^{k} D_{TT}^{\prime\perp}(z,k_{\perp}) \nonumber\\ 
 & \hspace{2cm} + k_{\perp\alpha} \Bigl[ S_{LL}D_{LL}^\perp(z,k_{\perp}) + \frac{k_{\perp} \cdot S_{LT}}{M}D_{LT}^\perp (z,k_{\perp}) 
 + \frac{S_{TT}^{kk}}{M^2} D_{TT}^\perp(z,k_{\perp}) \Bigr] \nonumber\\
 & \hspace{2cm} + \frac{M^2}{p^+}n_\alpha \Bigl[ S_{LL} D_{3LL}(z,k_{\perp}) + \frac{k_{\perp} \cdot S_{LT}}{M}D_{3LT}^\perp (z,k_{\perp})
 + \frac{S_{TT}^{kk}}{M^2} D_{3TT}^\perp(z,k_{\perp}) \Bigr],  \label{eq:XiTV}\\
& z\tilde \Xi_\alpha^{T(0)}(z,k_{\perp};p,S) = p^+ \bar n_\alpha \Bigl[ \frac{\tilde k_{\perp} \cdot S_{LT}}{M}G_{1LT}^\perp(z,k_{\perp}) 
  + \frac{S_{TT}^{\tilde kk}}{M^2}G_{1TT}^\perp(z,k_{\perp}) \Bigr] \nonumber\\
 & \hspace{2cm} - M\tilde S_{LT\alpha} G_{LT}(z,k_{\perp}) 
 - \tilde S_{TT\alpha}^{k} G_{TT}^{\prime\perp}(z,k_{\perp}) \nonumber\\
& \hspace{2cm} - \tilde k_{\perp\alpha} \Bigl[ S_{LL}G_{LL}^\perp(z,k_{\perp})  
  + \frac{k_{\perp} \cdot S_{LT}}{M}G_{LT}^\perp (z,k_{\perp}) + \frac{S_{TT}^{kk}}{M^2} G_{TT}^\perp(z,k_{\perp}) \Bigr] \nonumber\\
 & \hspace{2cm} + \frac{M^2}{p^+}n_\alpha \Bigl[ \frac{ \tilde k_{\perp} \cdot S_{LT} }{M}G_{3LT}^\perp(z,k_{\perp}) 
 + \frac{S_{TT}^{\tilde kk}}{M^2} G_{3TT}^\perp(z,k_{\perp}) \Bigr]  , \label{eq:XiTAV}\\
& z\Xi_{\rho\alpha}^{T(0)}(z,k_{\perp} ;p,S) = -p^+ \bar n_{[\rho} \tilde S_{LT\alpha]} H_{1LT}(z,k_{\perp}) 
 - \frac{p^+}{M} \bar n_{[\rho} \tilde S_{TT\alpha]}^{k} H_{1TT}^{\prime\perp}(z,k_{\perp}) \nonumber\\
 & \hspace{2cm} -\frac{p^+}{M} \bar n_{[\rho} \tilde k_{\perp\alpha]} \Bigl[ S_{LL} H_{1LL}^\perp(z,k_{\perp}) 
+ \frac{k_{\perp} \cdot S_{LT}}{M}H_{1LT}^\perp (z,k_{\perp}) + \frac{S_{TT}^{kk}}{M^2} H_{1TT}^\perp(z,k_{\perp}) \Bigr] \nonumber\\
 & \hspace{2cm} + M\varepsilon_{\perp\rho\alpha} \Bigl[ S_{LL} H_{LL}(z,k_{\perp}) + \frac{k_{\perp} \cdot S_{LT}}{M}H_{LT}^\perp (z,k_{\perp}) 
 + \frac{S_{TT}^{kk}}{M^2} H_{TT}^\perp(z,k_{\perp}) \Bigr] \nonumber\\
 & \hspace{2cm} + \bar n_{[\rho}n_{\alpha]} \Bigl[ (\tilde k_{\perp} \cdot S_{LT}) H_{LT}^{\prime\perp}(z,k_{\perp}) 
 + \frac{S_{TT}^{\tilde kk}}{M}H_{TT}^{\prime\perp}(z,k_{\perp}) \Bigr] \nonumber\\
 & \hspace{2cm} - \frac{M}{p^+} n_{[\rho} \tilde k_{\perp\alpha]} \Bigl[ S_{LL} H_{3LL}^\perp(z,k_{\perp}) 
 + \frac{k_{\perp} \cdot S_{LT}}{M}H_{3LT}^\perp (z,k_{\perp}) + \frac{S_{TT}^{kk}}{M^2} H_{3TT}^\perp(z,k_{\perp}) \Bigr] \nonumber\\
 & \hspace{2cm} - \frac{M}{p^+} n_{[\rho M \tilde S_{LT\alpha}]}\Bigl[  H_{3LT}(z,k_{\perp}) 
 + \tilde S_{TT\alpha]}^{k}H_{3TT}^{\prime\perp}(z,k_{\perp}) \Bigr].  \label{eq:XiTT}
\end{align}
\end{adjustwidth}

We have used the shorthanded notations such as $S_{TT}^{kk} \equiv S_{TT}^{\alpha\beta} k_{\perp\alpha}k_{\perp\beta}$. There are in total 40 tensor polarization-dependent TMD FFs; of these.10 contribute at the leading twist, 20 contribute at twist-3, and the remaining 10 contribute at twist-4. 
The 24 TMD FFs defined from the decomposition of $\tilde \Xi_\alpha^{T(0)}$ and $\Xi_{\rho\alpha}^{T(0)}$ are naive T-odd.
Among these TMD FFs, we notice in particularly that the $S_{LL}$ dependent TMD FF $D_{1LL}$, which is responsible for the spin alignment of the produced vector meson, is decoupled from the quark polarization. This suggests that the vector meson spin alignment can also be observed in the unpolarized high-energy collisions~\cite{Chen:2016iey,Chen:2020pty,STAR:2022fan}. Besides, $D_{1LL}$ also survives the $k_\perp$-integral. Therefore, it also appears in the colinear factorization. 

The rank-3-polarization-tensor-dependent TMD FFs are unique for spin-3/2 (or higher) hadrons.
A complete set of leading twist quark TMD FFs for spin-3/2 hadrons has been given in~\cite{Zhao:2022lbw}.
There are in total 14 rank-3-polarization-tensor-dependent TMD FFs that can be defined at the leading twist level.
We refer interested readers to~\cite{Zhao:2022lbw} for a detailed discussion.

\subsection{TMD FFs of Antiquarks and Gluons}
\label{sec:antiquark-ff}

One can define antiquark TMD FFs by replacing the fermion fields in the correlator of quark TMD FFs with the charge-conjugated fields.
Therefore, it is easy to find that the traces of the correlator with Dirac matrices $I$, $i\gamma_5$ and $\gamma^\mu \gamma_5$ will have an opposite sign between quark and antiquark cases, while the traces with $\gamma^\mu$ and $i\sigma^\mu\gamma_5$ are the same~\cite{Mulders:1995dh,Boer:1997mf,Pitonyak:2013dsu}.
The definition and parameterization of the antiquark TMD FFs are then full analogous to those of quark TMD FFs.

The gluon FFs are defined through the gluon correlator given by~\cite{Collins:1981uw,Mulders:2000sh}
\begin{align}
\hat \Gamma^{\mu\nu;\rho\sigma}(k;p,S)
= \sum_X \int \frac{d^4\xi}{(2\pi)^4} e^{ik\cdot \xi}
\langle 0\vert  F^{\rho\sigma}(\xi) \vert p,S; X\rangle 
\langle p,S;X\vert {\cal U}(\xi,0) F^{\mu\nu}(0)\vert 0\rangle,
\end{align}
where $F^{\rho\sigma}(\xi)\equiv F^{\rho\sigma,a}T^a$ is the gluon field field strength tensor, and ${\cal U}(\xi,0)$ is the Wilson line in the adjoint representation that renders the correlator gauge invariant. Under the assumption that the fragmenting parton moves in the plus direction, an integration over the $k^-$ component is carried out to give the TMD gluon correlator.

At the leading twist, we need to consider
\begin{align}
M \hat \Gamma^{ij}(z,k_\perp;p,S) =
\int dk^- \ \Gamma^{+j;+i}(k;p,S),
\end{align}
where $i$ and $j$ are transverse Lorentz indices in the transverse directions.

For the spin-1/2 hadron production, there are eight leading twist gluon TMD FFs which are given by the decomposition of the TMD gluon correlator~\cite{Mulders:2000sh}. We have the following:
\begin{align}
\hat{\Gamma}_U^{i j} (z, k_\perp; p, S) =
& \frac{p^{+}}{M}\left[-g_T^{i j} D_{1g} (z, k_\perp)+\left(\frac{k_\perp^i k_\perp^j}{M^2}+g_T^{i j} \frac{\boldsymbol{k}_\perp^2}{2 M^2}\right) H_{1g}^{\perp}(z, k_\perp )\right], \nonumber\\
\hat{\Gamma}_L^{i j}(z, k_\perp; p, S) =
&  - \lambda \frac{p^{+}}{M}\left[ i \varepsilon_\perp^{i j} G_{1Lg} (z, k_\perp)-\frac{\varepsilon_\perp^{k_\perp\{i} k_\perp^{j\}}}{2 M^2} H_{1Lg}^{\perp}(z, k_\perp)\right], \nonumber\\
\hat{\Gamma}_T^{i j}(z, k_\perp; p, S) =
& - \frac{p^{+}}{M}\Biggl[ g_T^{i j} \frac{\varepsilon_\perp^{k_\perp S_T}}{M} D_{1Tg}^\perp (z, k_\perp) + i \varepsilon_\perp^{i j} \frac{\boldsymbol{k}_\perp \cdot \boldsymbol{S}_{T}}{M} G_{1Tg}^{\perp} (z, k_\perp) \nonumber \\
& 
- \frac{\varepsilon_\perp^{k_\perp\{i} S_T^{j\}}+\varepsilon_\perp^{S_T\{i} k_\perp^{j\}}}{4 M}  H_{1Tg} (z, k_\perp) 
- \frac{\varepsilon_\perp^{k_\perp\{i} k_\perp^{j\}}}{2 M^2} \frac{\boldsymbol{k}_\perp \cdot \boldsymbol{S}_{T}}{M} H_{1Tg}^{\perp}(z, k_\perp) \Biggr].
\end{align}

$\hat \Gamma_U$, $\hat \Gamma_L$, and $\hat \Gamma_T$ stand for the unpolarized, the longitudinal, and transverse polarized parts for the hadron production, respectively. Analogously to the quark FFs, we have used $D$ to represent FFs of the unpolarized gluons, $G$ to represent the FFs of the circularly polarized gluons, and $H$ to represent the FFs of the linearly polarized gluons. Higher twist gluon TMD FFs are also discussed in ~\cite{Mulders:2000sh}, who further detail the parameterizations.

\section{Experiment and Phenomenology}
\label{sec:exp}

In high-energy experiments, the polarization of final state hadrons is usually measured from the angular distribution of their decay products. It is very challenging to acquire accurate experimental data. In light of a considerably large amount of free parameters, the spin-dependent FFs are not well-constrained experimentally. Compared with the case for unpolarized PDFs or FFs, the quantitative study of spin-dependent FFs is still immature. That said, there are already quite a few phenomenological studies making full use of the available experimental data. In this section, we summarize the available experimental data and the corresponding phenomenological studies.

\subsection{$\Lambda$ Hyperons}

The polarization of $\Lambda^0$ hyperons is usually measured from the angular distribution of the daughter proton in the parity-violating $\Lambda^0 \to p + \pi^-$ decay channel. In the rest frame of $\Lambda^0$, the normalized angular distribution of the daughter proton reads as follows:
\begin{align}
\frac{1}{N} \frac{dN}{d\cos\theta^*} = \frac{1}{2} (1 + \alpha {\cal P} \cos\theta^*) ,
\end{align}
where $\alpha = 0.732 \pm 0.014$ is the decay parameter of $\Lambda$ \cite{ParticleDataGroup:2020ssz}, ${\cal P}$ is the polarization of $\Lambda$ along a specified direction, and $\theta^*$ is the angle between the proton momentum and the specified direction to measure the $\Lambda$ polarization. 

The LEP experiment is an $e^+e^-$ collider at the $Z^0$-pole. Due to the parity violation in the weak interaction, the produced quark and antiquark are strongly polarized along the longitudinal direction. The longitudinal polarizations of those final state quarks and antiquarks in $e^+e^-$ annihilation at different collisional energies can be easily computed at the LO level and are explicitly shown in~\cite{Augustin:1978wf}. At the $Z^0$-pole, the longitudinal polarization of the final state down-type quarks can reach $0.9$. That of the up-type quarks is a bit smaller but is still about $0.6 \sim 0.7$. Based on the $SU(6)$ spin-flavor symmetry, the polarization of $\Lambda^0$ is determined by the polarization of the $s$ quark. It is thus proposed in~\cite{Gustafson:1992iq} that the final state $\Lambda^0$ hyperons are also strongly polarized at LEP, and the measurement of this polarization can probe interesting information on the hadronization mechanism. In the language of QCD factorization, the LEP experiment is the ideal place to study the longitudinal spin transfer $G_{1L} (z)$, which represents the number density of producing longitudinally polarized $\Lambda^0$ hyperons from longitudinally polarized quarks. It is the $p_T$-integrated version of the TMD FF $G_{1L}(z,p_\perp)$. 

At the leading order and leading twist, the longitudinal polarization of $\Lambda^0$ \linebreak reads as follows: \cite{Augustin:1978wf, Chen:2016iey}
\begin{align}
{\cal P}_L (y,z) = \frac{\sum_q \lambda_q (y) \omega_q (y) G_{1L,q} (z) + \{ q \leftrightarrow \bar q; y \leftrightarrow (1-y) \} }{\sum_q \omega_q (y) D_{1,q} (z) + \{ q \leftrightarrow \bar q; y \leftrightarrow (1-y) \}},
\end{align}
where $\lambda_q (y) = \Delta \omega_q (y) / \omega_q (y)$ is the helicity of the fragmenting quark with $\Delta \omega_q (y)$ and $\omega_q (y)$ being defined as follows:
\begin{align}
&
\Delta \omega_q (y) = \chi T_1^q (y) + \chi_{\rm int}^q I_1^q (y),
\\
&
\omega_q (y) = \chi T_0^q (y) + \chi_{\rm int}^q I_0^q (y) + e_q^2 A(y),
\\
&
T_1^q(y) = - 2c_V^q c_A^q [(c_V^e)^2 + (c_A^e)^2] A(y) + 2[(c_V^q)^2 + (c_A^q)^2] c_V^e c_A^e B (y),
\\
&
T_0^q (y) = [(c_V^q)^2 + (c_A^q)^2] [(c_V^e)^2 + (c_A^e)^2] A(y) - 4 c_V^q c_A^q c_V^e c_A^e B(y),
\\
&
I_1^q (y) = - c_A^q c_V^e A(y) + c_V^q c_A^e B(y),
\\
&
I_0^q (y) = c_V^q c_V^e A(y) - c_A^q c_A^e B(y).
\end{align}

Here, $y=(1+\cos\theta)/2$ with $\theta$ is the angle between the outgoing $\Lambda$ and the incoming electron. The coefficient functions are given as $A(y)= (1-y)^2 + y^2$, $B(y) = 1-2y$, $\chi = Q^4/[(Q^2-M_Z^2)^2 + \Gamma_Z^2 M_Z^2] \sin^4 2\theta_W$, and $\chi_{\rm int}^q = -2 e_q Q^2(Q^2-M_Z^2)/[(Q^2-M_Z^2)^2 + \Gamma_Z^2 M_Z^2] \sin^2 2\theta_W$. $c_V^{q/e}$ and $c_A^{q/e}$ are the coupling constants of the vector current and axis-vector current parts of the quark/electron, with $Z^0$. $M_Z$ being the mass of $Z^0$ and $\Gamma_Z$ being the width. Notice that $\lambda_{\bar q} (y) = - \lambda_q (1-y)$. The quark helicity and antiquark helicity have the opposite sign. This is in line with the sign flip in Section~\ref{sec:antiquark-ff}. Therefore, the polarization of $\bar\Lambda^0$ is expected to have the opposite sign with that of $\Lambda^0$. 

\begin{figure} [H]
    \includegraphics[width=0.60\textwidth]{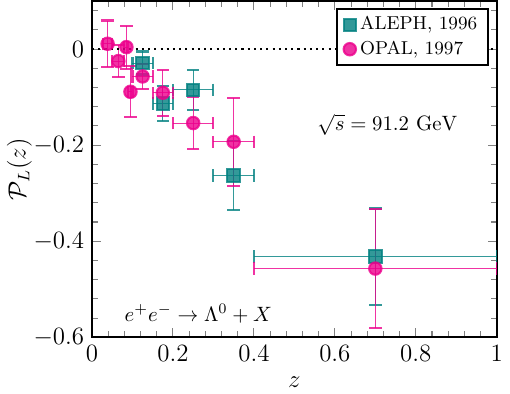}
    \caption{Reproduction of the longitudinal polarization of $\Lambda^0$ in $e^+e^-$ annihilation at $\sqrt{s}=91.2$ GeV measured by the ALEPH \cite{ALEPH:1996oew} and OPAL \cite{OPAL:1997oem} collaborations at LEP. We have combined the statistical and systematic errors. Neglecting the mass of $\Lambda$ hyperons in the high-energy limit, the definitions of $z$ in these two experiments are the same as those of the momentum fraction in the light-cone coordinate currently used in the QCD factorizations.} 
    \label{fig:lep-lambda}
\end{figure}

Since the quark helicity $\lambda_q (y)$ and the production weight $\omega_q (y)$ are calculable in quantum field theory, the measurement of the longitudinal polarization of final state $\Lambda^0$ as a function of $z$ can directly provide information of the longitudinal spin transfer. Such experiments were eventually carried out by ALEPH and OPAL collaborations at LEP in the 1990s \cite{ALEPH:1996oew, OPAL:1997oem}. As shown in Figure~\ref{fig:lep-lambda}, the longitudinal polarization increases monotonically with increasing $z$, which provides a hint on how to parameterize the longitudinal \linebreak spin transfer. 

Following the release of these experimental data, many phenomenological \linebreak studies \cite{Kotzinian:1997vd, deFlorian:1997zj, Liang:1997rt, Boros:1998kc,Ma:1998pd,Ma:1999gj,Ma:1999wp,Liu:2000fi,Chen:2016iey} were carried out to understand the longitudinal spin \textls[-25]{transfer $G_{1L}(z)$. Among them, the de Florian--Stratmann--Vogelsang (DSV) \mbox{parameterization \cite{deFlorian:1997zj}}} offers three scenarios. The first scenario is based on the naive parton model, which assumes that only the $s$ quark contributes to the longitudinal spin transfer at the initial scale. The second scenario assumes that the $u$ and $d$ quarks contribute to negative $G_{1L}(z)$ at the initial scale. The third scenario assumes that $u$, $d$, and $s$ contribute equally. All three can describe the experimental data reasonably well. A more recent Chen--Yang--Zhou--Liang (CYZL) \mbox{analysis \cite{Chen:2016iey}} also obtained a good description of the experimental data utilizing the LO formula. The ambiguity again highlights the difficulties in the quantitative study of FFs. It can only be removed through a global analysis of the experimental data in various high-energy reactions. Therefore, many works have also made predictions for the longitudinal polarization of $\Lambda$ produced in polarized SIDIS \cite{Jaffe:1996wp, Kotzinian:1997vd, deFlorian:1997kt, Ashery:1999am} and pp collisions \cite{deFlorian:1998ba, Jager:2002xm, Xu:2002hz, Xu:2005ru}. 

\begin{figure}[H]
\begin{adjustwidth}{-\extralength}{0cm}
\centering 
\includegraphics[width=0.60\textwidth]{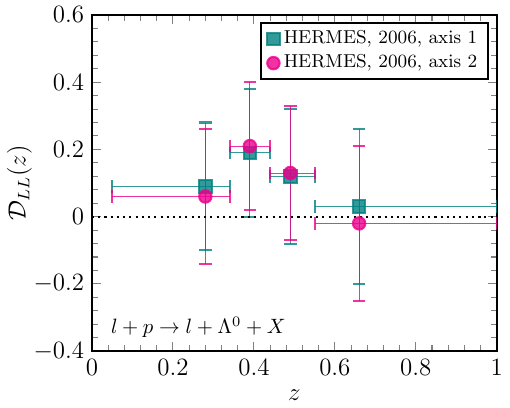}
\includegraphics[width=0.60\textwidth]{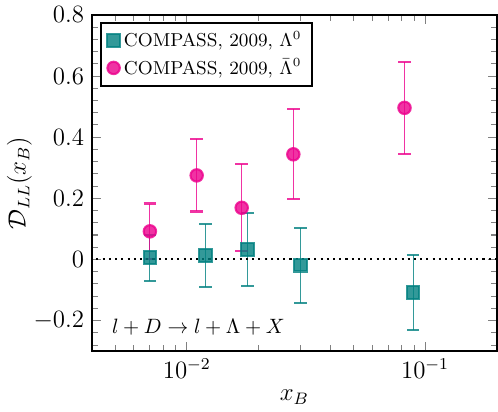}
\end{adjustwidth}
\caption{Spin transfer coefficient ${\cal D}_{LL}$ as a function of $z$ and $x_B$ in polarized SIDIS measured by the HERMES \cite{HERMES:1999buc} and COMPASS \cite{COMPASS:2009nhs} collaborations. Only the statistic errors are shown in both plots. Axes 1 and 2 refer to two different definitions of the longitudinal direction of $\Lambda$ in the experiment. They are approximately the same at the high-energy limit. However, at the HERMES energy, they are not parallel to each other.}
\label{fig:sidis-data}
\end{figure}

The inclusive DIS process with the polarized lepton beam has been used to probe the spin structure of the nucleon \cite{SpinMuon:1998eqa, E143:1998hbs, HERMES:1998cbu}. In this process, only the momentum of the final state lepton was measured. Therefore, we can gain information on the nucleon structure but lose those on the hadronization. To restore the access to (spin-dependent) FFs, we have to rely on the semi-inclusive process and measure (the polarization of) one final state hadron (There are two fragmentation regimes in SIDIS, namely the current fragmentation and the target fragmentation. Although the target fragmentation function is also currently a hot topic, it is beyond the scope of this review. We only focus on the study of the current fragmentation function). However, it is not a simple task to do so in the real world. Despite the difficulties, early attempts from the E665 \cite{E665:1999fso} and HERMES \cite{HERMES:1999buc} collaborations were still successfully performed. Recent measurements from HERMES \cite{HERMES:2006lro} and COMPASS \cite{COMPASS:2009nhs} collaborations have also elevated the quality of experimental data to a level that sheds light on phenomenological studies. These experiments measure the spin transfer coefficient ${\cal D}_{LL}(z)$ (it is important to not get confused with the spin-alignment-dependent FF $D_{1LL}(z)$ of vector mesons) which, at the leading order and leading twist approximation, is given by \cite{Mulders:1995dh}
\begin{align}
{\cal D}_{LL} (x_B,z) = \frac{\sum_q e_q^2 x_B f_{1,q} (x_B) G_{1,q} (z)}{\sum_q e_q^2 x_B f_{1,q} (x_B) D_{1,q} (z)},
\end{align}
with $f_{1,q} (x_B)$ being the unpolarized PDF. Due to the presence of the unpolarized PDF of proton/nucleus, the polarized SIDIS experiment favors more contributions from the $u$ and $d$ quarks at large $x_B$ than from the $e^+e^-$ collider. We show the HERMES data set as a function of $z$ (integrating over $x_B$) and the COMPASS data set as a function of $x_B$ (integrating over $z$) \linebreak in Figure~\ref{fig:sidis-data}. 

The experimental data from E665 \cite{E665:1999fso} suggested a difference between ${\cal D}_{LL}$ for the $\Lambda^0$ production and that for the $\bar\Lambda^0$ production. This was later confirmed by the \linebreak COMPASS \cite{COMPASS:2009nhs} experiment. In~\cite{Ma:2000uu,Zhou:2009mx,Chi:2014xba}, it was shown that such a difference serves as a flavor tag in the study of the $G_{1L}$ FF. More studies on the flavor dependence of PDFs/FFs have been performed~\cite{Ma:1999gj, Gluck:2000dy, Ma:2000ip, Ma:2000cg, Yang:2001yda, Leader:2001kh, Blumlein:2002qeu, Leader:2002az, Leader:2005ci}.  The NOMAD collaboration also carried out similar measurements in the neutrino SIDIS experiment \cite{NOMAD:2000wdf, NOMAD:2001iup}. Because of the flavor-changing feature of the charged weak interaction, this experiment opens more opportunities for quantitative research on the flavor dependence of spin-dependent FFs. A sophisticated investigation was presented in~\cite{Ellis:2002zv}. 

RHIC is the first and, so far, the only polarized proton--proton collider. The helicity of the incident protons can be transferred to that of the partons through the longitudinal spin transfer $g_{1L}(x)$ of PDFs. Therefore, it also has the capability of probing $G_{1L}(z)$ of the fragmentation. The first measurement was performed in 2009 \cite{STAR:2009hex}, while an improved analysis was presented in 2018 \cite{STAR:2018pps}. These experiments measure the spin transfer coefficient ${\cal D}_{LL}$ which is defined as follows:
\begin{align}
{\cal D}_{LL} \equiv \frac{\sigma^{p^+ p \to \Lambda^+ + X} - \sigma^{p^+ p \to \Lambda^- + X}}{\sigma^{p^+ p \to \Lambda^+ + X} + \sigma^{p^+ p \to \Lambda^- + X}}.
\end{align}

\begin{figure}[H]
    \includegraphics[width=0.60\textwidth]{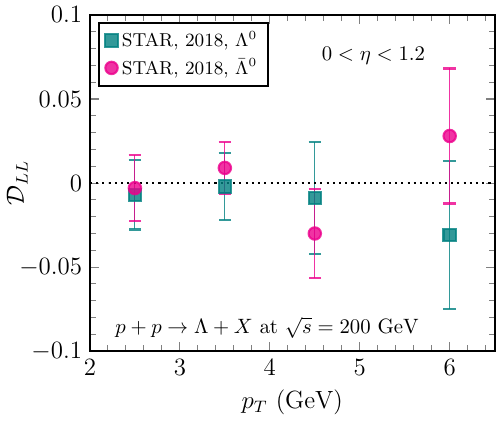}
    \caption{The spin transfer coefficient ${\cal D}_{LL}$ in polarized $pp$ collisions at $\sqrt{s}=200$ GeV measured by the STAR collaboration at RHIC \cite{STAR:2018pps}. Data points are taken from~\cite{STAR:2018pps}. The systematic error and the statistical error have been combined.}
    \label{fig:star-2018-data}
\end{figure}

The $+$ symbol in the superscript denotes the helicity of the corresponding proton or $\Lambda$ hyperon. The updated experimental data from the STAR collaboration \cite{STAR:2018pps} at RHIC are shown in Figure~\ref{fig:star-2018-data}. This experimental data tend to favor the first and second scenarios in the DSV parameterization \cite{deFlorian:1997zj}. However, it cannot concretely rule out any scenario yet due to the large uncertainties. Moreover, the Xu--Liang--Sichtermann approach \cite{Xu:2005ru} based on the $SU(6)$ spin-flavor symmetry can also describe this data well. Moreover, RHIC also measured the transverse spin transfer coefficient ${\cal D}_{TT}$, which is sensitive to the convolution of the transversity PDF and the transversity FF \cite{deFlorian:1998am}.

The polarizations of partons participating the same hard scattering are strongly correlated. The helicity amplitudes of different partonic processes have been evaluated and summarized in~\cite{Gastmans:1990xh}. Thus,~\cite{Chen:1994ar} proposes the dihadron polarization correlation as a probe to the longitudinal spin transfer $G_{1L}$ in $e^+e^-$ annihilations at low energy where the fragmenting quarks are not polarized. Recently, this idea was further investigated and applied to the unpolarized $pp$ collisions in~\cite{Zhang:2023ugf}. By measuring the longitudinal polarization correlation of two almost back-to-back hadrons, we also gain access to the longitudinal spin transfer in unpolarized $pp$ collisions. Since this observable avoids the contamination from the longitudinal spin transfer $g_{1L}$ in PDFs. which is also poorly known,~\cite{Zhang:2023ugf} innovated a means to investigating the longitudinal spin transfer $G_{1L}$ in FFs at RHIC, Tevatron, and the LHC. Furthermore, this work can also be used to constrain the FF of circularly polarized gluons.

\begin{figure}[H]
\begin{adjustwidth}{-\extralength}{0cm}
\centering
\includegraphics[width=0.29\textwidth]{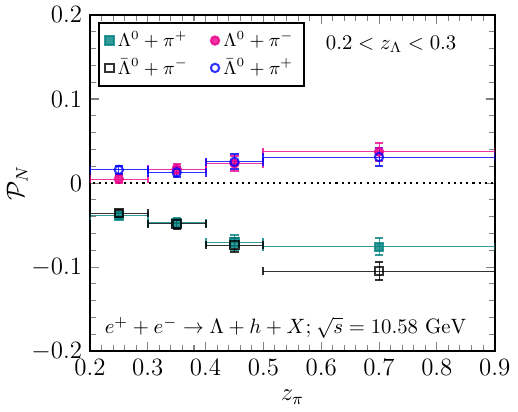}
\includegraphics[width=0.29\textwidth]{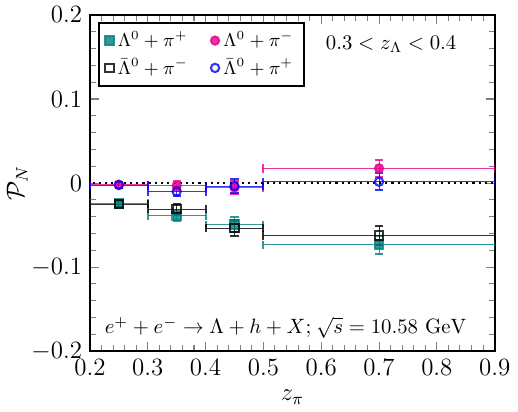}
\includegraphics[width=0.29\textwidth]{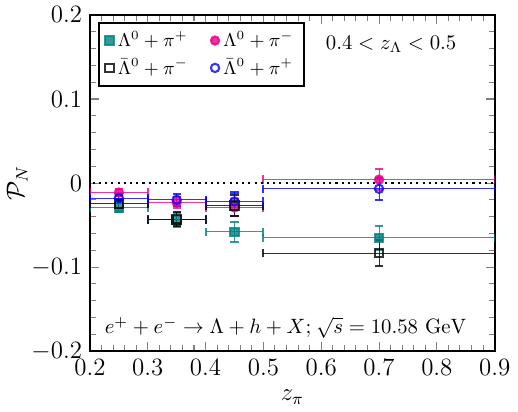}
\includegraphics[width=0.29\textwidth]{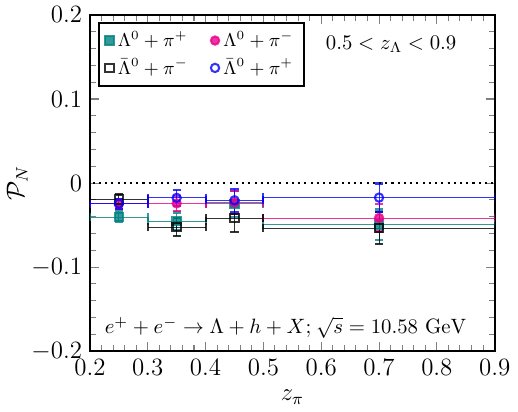}
\\
\includegraphics[width=0.29\textwidth]{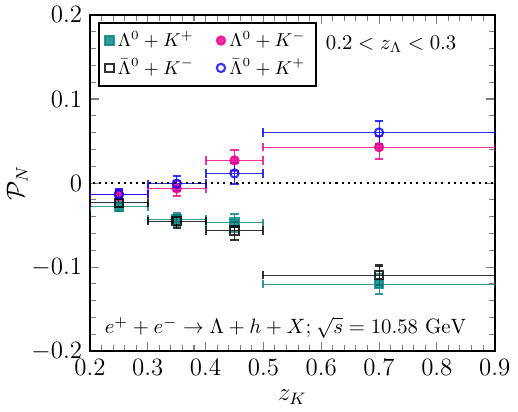}
\includegraphics[width=0.29\textwidth]{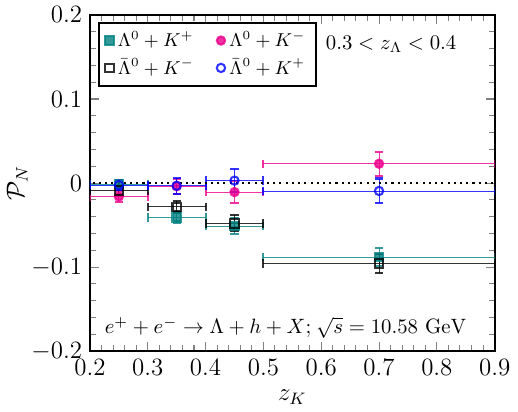}
\includegraphics[width=0.29\textwidth]{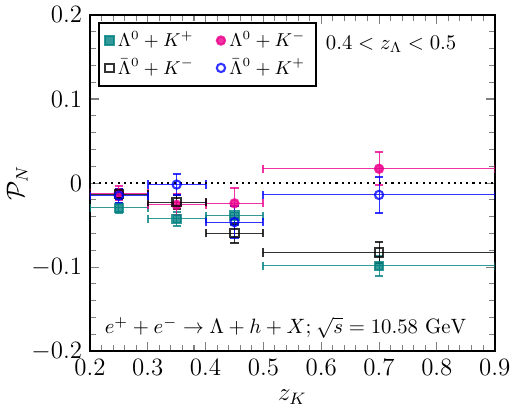}
\includegraphics[width=0.29\textwidth]{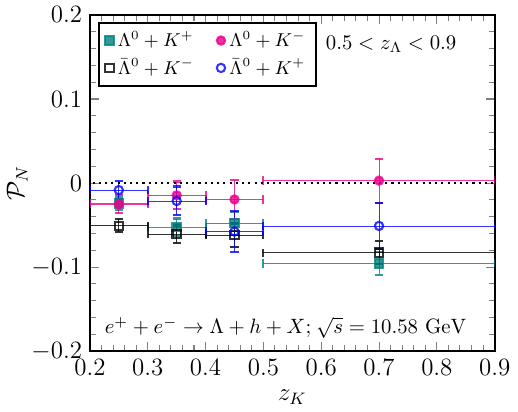}
\end{adjustwidth}
\caption{Transverse polarization of $\Lambda$ in $e^+e^-$ annihilation measured by the Belle collaboration \cite{Belle:2018ttu}. Data points are taken from~\cite{Belle:2018ttu}. Statistical and systematic errors are combined in quadrature.}
\label{fig:belle-pn}
\end{figure}

Recently, the Belle collaboration measured the transverse polarization of $\Lambda$ hyperons in $e^+e^-$ annihilations \cite{Belle:2018ttu}, sparking considerable theoretical interest \cite{Matevosyan:2018jht, Gamberg:2018fwy, Anselmino:2019cqd, Anselmino:2020vlp, DAlesio:2020wjq, Callos:2020qtu, Kang:2020xyq, Boglione:2020cwn, Li:2020oto, Chen:2021hdn, Gamberg:2021iat, DAlesio:2021dcx, DAlesio:2022brl,Boglione:2022nzq}. In this experiment, one first defines the hadron production plane and then measures the transverse polarization along its normal direction. Since there are two transverse directions, we refer to the polarization along one as ${\cal P}_N$ and the other one as ${\cal P}_T$. The hadron production plane can be defined in two ways. The first one is defined by the thrust axis and the $\Lambda$ momentum. In the second, the thrust axis is replaced by the momentum of a reference hadron (in the back-to-back side). Therefore, this experiment is dedicated to probing the $D_{1T}^\perp (z,p_\perp)$ FF. While the $p_T$-differential experimental data of ${\cal P}_{N}$ contain sizable uncertainties, the $p_T$-integrated version is quite precise, as shown in Figure~\ref{fig:belle-pn}. Employing the Trento convention \cite{Bacchetta:2004jz} for the definition of $D_{1T}^\perp (z,p_\perp)$, the $p_\perp$ integrated transverse polarization is given by \cite{DAlesio:2020wjq, Callos:2020qtu,Chen:2021hdn,Chen:2021zrr}
\begin{align}
{\cal P}_{N} (z_\Lambda) = \frac{\sum_q e_q^2 \int d^2 \bm{p}_\perp d^2 \bm{p}_{h\perp} \frac{- \hat P_{\perp\Lambda} \cdot p_\perp}{z_\Lambda M_\Lambda} D_{1,q}^h (z_h,p_{h\perp}) D_{1T,q}^{\perp\Lambda} (z_\Lambda,p_{\perp})}{\sum_q e_q^2 \int d^2 \bm{p}_\perp d^2 \bm{p}_{h\perp} D_{1,q}^h (z_h,p_{h\perp}) D_{1,q}^{\Lambda} (z_\Lambda,p_{\perp})} \Biggr|_{\bm{P}_{\perp \Lambda} = \frac{z_\Lambda}{z_h} \bm{p}_{h\perp} + \bm{p}_\perp },
\end{align}
where $\hat P_{\perp\Lambda}$ is the unit vector along the direction of $P_{\perp\Lambda}$. The integral in the denominator simply reduces to the product of two colinear FFs. However, to evaluate the numerator, we need to first parameterize the $p_\perp$ and $p_{h\perp}$ dependence at the initial scale, which then evolves to the TMD factorization scale through use of the Collins--Soper--Sterman evolution equation. Nonetheless, since the collisional energy at Belle is not very high, a Gaussian ansatz is already a good approximation. More sophisticated approaches incorporating the $p_\perp$ dependence can be found in~\cite{Gamberg:2021iat, DAlesio:2022brl,Boglione:2022nzq}.

As shown in Figure~\ref{fig:belle-pn}, the distinct difference between ${\cal P}_{N}$ measured in the $\Lambda+\pi^+({\rm or~} K^+)$ and $\Lambda+\pi^-({\rm or~} K^-)$ processes offers an opportunity to explore the flavor dependence. Early attempts, such as the D'Alesio--Murgia--Zaccheddu (DMZ) \cite{DAlesio:2020wjq} and Callos-Kang--Terry (CKT) \cite{Callos:2020qtu} parameterizations, adopted the strategy that valence parton FFs differ from each other and that parton FFs are the same, i.e., $D_{1T,u}^{\perp \Lambda} \neq D_{1T,d}^{\perp \Lambda} \neq D_{1T,s}^{\perp \Lambda} \neq D_{1T,{\rm sea}}^{\perp \Lambda}$. However, this approach violates the isospin symmetry, which is one of the most important features of strong interaction. Furthermore, a model calculation \cite{Li:2020oto} based on the strict $SU(6)$ spin-flavor symmetry failed to describe the experimental data. However, it was first shown in~\cite{Chen:2021hdn} that the isospin symmetric Chen--Liang--Pan--Song--Wei (CLPSW) parameterization can still describe the experimental data well as long as the artificial constraint on sea parton FFs is released. This perspective was further investigated in Ref.~\cite{DAlesio:2022brl} recently, which concluded that one can obtain good fit to the Belle data with and without implementing the isospin symmetry constraint after taking into account the charm contribution. This confirms that the current Belle dataset does not represent an isospin symmetry violation in the hadronization. Furthermore,~\cite{Chen:2021zrr} proposed to test the isospin symmetry at the future EIC experiment. By comparing the transverse polarizations in $ep$ and $eA$ scatterings at large $x$, we can ultimately check the difference between $D_{1T,u}^\perp$ and $D_{1T,d}^\perp$. 

The future EIC is a polarized electron--proton/ion collider with unprecedentedly high luminosity. It will open a new window for the quantitative study of spin-dependent FFs. Several works \cite{Yang:2021zgy, Li:2021txj, Kang:2021ffh, Kang:2021kpt} have proposed and made predictions for different observables at the future EIC with polarized proton beams. These observables are sensitive to various combinations of spin-dependent PDFs and FFs. Therefore, the future measurement will reveal information on both hadron structure and hadronization. A recent work \cite{Chen:2021zrr} also proposed a method to study spin-dependent PDFs/FFs in unpolarized experiments. The key idea is that the polarizations of the final state quark and initial state parton are correlated. Thanks to the Boer--Mulders function in the PDFs, the initial state quark are transversely polarized although the polarization depends on the azimuthal angle. This transverse polarization can further propagate into final state observables through chiral-odd FFs. By measuring the azimuthal-angle-dependent longitudinal and transverse polarizations of final state $\Lambda$, we can probe $H_{1L}^\perp$ and $H_{1T}^\perp$ even in the unpolarized SIDIS process. Moreover, we can also measure the azimuthal-angle-dependent polarizations in $e^+e^-$ annihilations to probe combinations of the Collins function and spin-dependent chiral-odd FFs \cite{Chen:2021zrr}. This idea is akin to those explored in~\cite{Ellis:2011kq, DAlesio:2021dcx}.

\subsection{Vector Mesons}

Most vector mesons decay through parity-conserving strong interactions. Their polarization vector does not enter the angular distribution of the daughter hadrons. Therefore, it is not possible to measure their polarization vector. In contrast, the tensor polarization does play a role in the angular distribution and therefore can be measured. 
Among them, spin alignment, which quantifies the deviation from 1/3 of $\rho_{00}$ in the spin-density matrix, has received the most attention.

\begin{figure}[H]
    \includegraphics[width=0.60\textwidth]{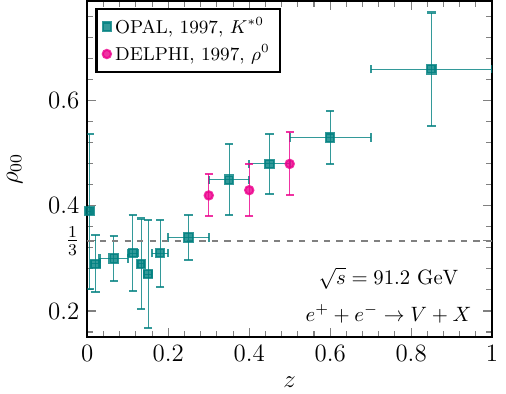}
    \caption{Spin alignment of $K^{*0}$ and $\rho^0$ measured by the OPAL \cite{OPAL:1997vmw} and DELPHI \cite{DELPHI:1997ruo} collaborations at LEP. Data points are taken from~ \cite{OPAL:1997vmw, DELPHI:1997ruo}.}
    \label{fig:lep-spin-alignment}
\end{figure}

Several collaborations \cite{DELPHI:1997ruo,OPAL:1997vmw,OPAL:1997nwj,OPAL:1999hxs} at LEP have measured the spin alignment of different vector mesons produced in the $e^+e^-$ annihilation at the $Z^0$-pole. We show the spin alignment of $K^{*0}$ and $\rho^0$ measured by the OPAL \cite{OPAL:1997vmw} and DELPHI \cite{DELPHI:1997ruo} collaborations in Figure~\ref{fig:lep-spin-alignment}. The off-diagonal matrix elements were also measured in some of the experiments. Thereafter, the NOMAD collaboration measured the vector meson spin alignment for the first time in the neutrino DIS experiment \cite{NOMAD:2006kuc}. These measurements offer more information on the hadronization mechanism and have led to several phenomenological studies \cite{Anselmino:1998jv, Anselmino:1999cg, Schafer:1999am, Xu:2001hz, Bacchetta:2001rb, Shlyapnikov:2001jf, Xu:2002vz, Xu:2003fq, Xu:2003rs}. 

Figure~\ref{fig:lep-spin-alignment} shows that $\rho_{00}$ is consistent with $1/3$ (i.e., no spin alignment) at the small-$z$ region. However, at large $z$, a clear spin alignment is observed. This pattern is similar to that for the longitudinal polarization of $\Lambda$ also measured at LEP \cite{ALEPH:1996oew, OPAL:1997oem} (shown \linebreak  in Figure~\ref{fig:lep-lambda}). 

As mentioned above, the quarks produced at LEP and also those at NOMAD are strongly polarized. Therefore, it is tempting to attribute the tensor polarization of final state vector mesons to the longitudinal polarization of the fragmenting quarks. However, a simple tensor structure analysis ~\cite{Bacchetta:2000jk, Bacchetta:2001rb, Wei:2013csa, Wei:2014pma} shows that this is not the case. The spin alignment of the final state mesons is not coupled with the quark polarization. Instead, it is coupled with the quark-polarization-summed cross section. The vector meson spin alignment in $e^+e^-$ collisions is given by
\begin{align}
\rho_{00} = \frac{1}{3} - \frac{1}{3} \frac{\sum_q \omega_q (y) D_{1LL,q} (z)}{\sum_q \omega_q (y) D_{1,q} (z)},
\end{align}
where, $\omega_q$ is defined to be the same as that for the $\Lambda$ production in the previous section, and $D_{1LL}(z)$ is the corresponding FF that is responsible for the vector meson spin alignment. As shown in the above equation, the longitudinal polarization of the fragmenting quark does not play a role here. It was thus first proposed in~\cite{Wei:2013csa} that the vector meson spin alignment can also be observed in other high-energy collisions with unpolarized quarks fragmenting. Fitting to the experimental data from LEP, other work~\cite{Chen:2016iey, Chen:2020pty} extracted $D_{1LL}(z)$ and made predictions for the spin alignment of high $p_T$ vector mesons in unpolarized pp collisions at RHIC and the LHC \cite{Chen:2020pty}. Furthermore, from the same mechanism, there will be a significant spin alignment for vector mesons produced in the unpolarized SIDIS. Measuring vector meson spin alignment at the future EIC will cast new light on the quantitative study of the $D_{1LL} (z)$ FF. 

Notice that the spin alignment of low-$p_T$ vector mesons in $AA$ collisions has also been measured at RHIC \cite{STAR:2008lcm,STAR:2022fan} and LHC \cite{Kundu:2021lra} recently. These low-$p_T$ hadrons in relativistic heavy-ion collisions are produced through a different hadronization mechanism than those of fragmentation. Their tensor polarization originates from a different source.

\section{Model Calculation}
\label{sec:model}

The PDFs and FFs are defined in terms of quark--gluon correlators as laid out in Section~\ref{sec:spin}. Owing to the nonperturbative nature of the hadron state, we cannot directly evaluate them theoretically. Thus far, several proposals for computing quantities that can be related to the PDFs in the lattice QCD approach have been put forward \cite{Ji:2013dva, Ji:2017oey, Radyushkin:2017cyf, Orginos:2017kos, HadStruc:2021wmh}. However, it is not possible to study FFs in the lattice QCD yet. In the current stage, the quantitative information is mainly extracted from the experimental data. 

However, due to the limited amount of experimental data, the TMD PDFs and FFs are not yet well constrained. As a complementary tool, model calculations have usually been employed to compute different PDFs over the past decades \cite{Chodos:1974je, Chodos:1974pn, Jakob:1997wg, Petrov:1998kf,Bentz:1999gx, Penttinen:1999th, Brodsky:2002cx, Brodsky:2002rv, Boer:2002ju, Brodsky:2002pr, Ji:2002aa, Gamberg:2003ey, Bacchetta:2003rz, Belitsky:2003nz, Yuan:2003wk, Lu:2004hu, Goeke:2006ef, Cloet:2007em, Meissner:2007rx, Bacchetta:2008af, Pasquini:2008ax, Avakian:2008dz, Avakian:2008mu, Boffi:2009sh, Efremov:2009ze, She:2009jq, Avakian:2010br, Pasquini:2010af, Lu:2016vqu, Ma:2019agv, Bacchetta:2020vty, Luan:2022fjc, Tan:2023kbl,Sharma:2023azu}. These investigations offer quantitative insight into the hadron structure and therefore are indispensable for phenomenological study. The same also goes for the FFs. Most of the models can be used to evaluate both PDFs and FFs. We make a nonexclusive brief summary on \linebreak FF calculations.

There are quite a few models that can be categorized as a spectator model \cite{Metz:2002iz, Gamberg:2008yt}. Among them, the quark--diquark model is a simple one which provides the quark--baryon--diantiquark vertex,so that the baryon FFs can be easily evaluated. In~\cite{Nzar:1995wb}, the colinear baryon FFs were calculated at the leading twist using the quark--diquark model, while in~\cite{Jakob:1997wg, Yang:2017cwi, Wang:2018wqo, Lu:2015wja, Yang:2016mxl}, the TMD FFs were further computed at the leading and subleading twists. To compute the meson FFs or the gluon FFs, we need an improved version which offers the vertex among the fragmenting parton, hadron, and spectator. In ~\cite{Bacchetta:2007wc}, the Collins function was calculated for pions and kaons in this method. Recently, in~\cite{Xie:2022lra}, approach to calculate leading twist gluon TMD FFs was presented. The chiral invariant model \cite{Manohar:1983md} investigated the chiral symmetry and the spontaneous breaking with an effective Lagrangian of quarks, gluons, and goldstone bosons. It can also be classified into the spectator model category. Utilizing this model, several authors~\cite{Ji:1993qx, Londergan:1996vf, Bacchetta:2002tk, Nam:2011hg, Nam:2012af,Yang:2013cza} calculated the pion and kaon FFs. In~\cite{Andrianov:1999zz}, an extended version was also developed to compute the vector meson FFs. Furthermore, several works \cite{Bacchetta:2001di, Bacchetta:2003xn, Gamberg:2003eg, Amrath:2005gv} have evaluated the FFs of different hadrons using a parameterized quark--hadron coupling.

\textls[-15]{The Nambu--Jona--Lasinio (NJL) model originates from~\cite{Nambu:1961tp, Nambu:1961fr} who developed an effective theory describing the quark--hadron interaction. It has been employed to evaluate PDFs of different hadrons \cite{Korpa:1989pp, Shigetani:1993dx, Davidson:1994uv, Bentz:1999gx, Davidson:2001cc, Nguyen:2011jy, Hutauruk:2016sug}. Incorporating with the Feynman--Field model (also known as the quark--jet model) established in~\cite{Field:1977fa, Field:1982dg}, the NJL=-jet model has been employed to calculate both colinear and TMD FFs of different hadrons \cite{Ito:2009zc, Matevosyan:2010hh, Matevosyan:2011ey, Matevosyan:2011vj, Matevosyan:2012ga, Casey:2012ux, Yang:2013cza, Bentz:2016rav}. Recent works have also computed FFs of gluon \cite{Yang:2016gnd} and charm quark \cite{Yang:2020wvt} with this approach. The Feynman--Field model relates the total FF to the first rank FF. However, it does not specify how to compute the first rank FF. Therefore, in principle, it can be hybridized with another model which provides with the first rank FF to prolong the applicability of the \linebreak corresponding model.}

We make a final remark on the model calculation to conclude this section. All the above-mentioned models compute FFs employing the effective Lagrangian of partons and hadrons of interest. While these calculations offer quantitative insight into the hadronization scheme, we should draw a line between conclusions that are model-dependent and \linebreak  those that are model-independent. 

\section{Summary} \label{6}

There are a multitude of topics within the subject of FFs. In this review, we constrain ourselves in a very limited scope that we are familiar with. First, we briefly summarized the derivation of the TMD factorization and the establishment of the QCD evolution equation at the leading twist level. The TMD factorization and the corresponding evolution at the higher twist level are still ongoing topics. Second, we are particularly interested in the spin-related effects. With the spin degree of freedom being taken into account, the interplay between the transverse momentum and the hadron/quark polarization presents a highly intriguing phenomena that can be investigated in experiments. As a result, we need to define more TMD FFs to fully describe the fragmentation process. In quantum field theory, TMD FFs are introduced in the decomposition of parton correlators. We summarized the final results up to the twist-4 level for spin-$0$, spin-$1/2$, and spin-$1$ hadron productions. Finally, although all the TMD FFs have clear definitions in terms of parton fields and hadron states, they are nonperturbative quantities that cannot be directly evaluated from quantum field theory. In contrast to TMD PDFs, FFs cannot be computed even in the lattice QCD approach. The quantitative investigation thus mainly concentrates on the extraction from experimental measurements and model calculations. We summarized several spin-related experiments conducted over the past decades and the corresponding phenomenological studies. In the last section, we also briefly presented several model calculations. 

The study of TMD FFs is still a very active field, and many mysteries remain to be explored. The Electron-Ion Collider (EIC) and the Electron-Ion Collider in China (EicC) have been proposed to be built as the new high-energy colliders in the next generation. They will provide new experimental data for the quantitative study of TMD FFs and can significantly boost our understanding of the hadronization mechanism.

\vspace{6pt}
\authorcontributions {All authors contribute to the writting and proofreading of this review.}

\funding {K.-B.C. is supported by the National Natural Science Foundation of China under grants no. 12005122 and no. 11947055, as well as by the Shandong Province Natural Science Foundation under grant no. ZR2020QA082. T.L. is supported in part by the National Natural Science Foundation of China under grants no. 12175117 and no. 20221017-1. Y.-K.S. is supported in part by the National Natural Science Foundation of China under grant no. 11505080, and by the Shandong Province Natural Science Foundation under grant no. ZR2018JL006. S.-Y.W. is supported by the Taishan Fellowship of Shandong Province for junior scientists.}

\dataavailability {Data available on request from the authors.}


\conflictsofinterest {The authors declare no conflict of interest.}

\clearpage
\begin{adjustwidth}{-\extralength}{0cm}

\reftitle{References}

\PublishersNote{}
\end{adjustwidth}
\end{document}